\documentclass[12pt]{article}

\usepackage{amsmath}
\usepackage[dvips]{graphicx}
\usepackage{bm}
\usepackage{color} 

\newcommand{\bea}{\begin{eqnarray}}  
\newcommand{\eea}{\end{eqnarray}}  

\baselineskip
\begin{document}

\begin{flushright}
%LTH 512\\
%MC-TH-01-06\\
\today\\
\end{flushright}

\begin{center}
\vspace*{0cm}

{\Large {\bf Colour dipoles and $\rho$,$\phi$ electroproduction}} \\

\vspace*{1cm}

J.R.~Forshaw$^{1}$, R.~Sandapen$^{1,2}$ and G.~Shaw$^{1}$

\vspace*{0.2cm}
$^{1}$Department of Physics and Astronomy,\\
University of Manchester,\\
Manchester. M13 9PL. England.

\vspace*{0.2cm}
$^{2}$Department of Physics, Engineering Physics and Optics\\
Laval University,\\
Quebec. G1K 7P4. Canada.
\end{center}
\vspace*{2cm}

\begin{abstract}
\noindent
We present a detailed comparison of a variety of predictions for diffractive
light vector meson production with the data collected at the HERA collider. 
All our calculations are performed within a dipole model framework and make use
of different models for the meson light-cone wavefunction. There are no
free parameters in any of the scenarios we consider. Generally we find
good agreement with the data using rather simple Gaussian motivated
wavefunctions in conjunction with dipole cross-sections which have 
been fitted to other data. 

\end{abstract}

\mbox{PACS Numbers: 12.40.Nn, 13.60.Le}

\newpage
\section{Introduction}

\noindent
In a previous paper, Forshaw, Kerley and Shaw (FKS) \cite{fks1}  
reported on a successful attempt to 
extract the cross-section for scattering
colour dipoles of fixed transverse size off protons using both
electroproduction and photoproduction $\gamma p$ total cross-section data, 
together with the constraint provided by the measured ratio
of the diffraction dissociation cross-section to the total cross-section
for real photons. Subsequently, the same model has been applied
to ``diffractive deep inelastic scatterring'' (DDIS)  
$\gamma^* p \rightarrow X  \,  p$  \cite{fks2} and also to deeply virtual 
Compton scattering (DVCS) 
$\gamma^* p \rightarrow \gamma \,  p$  \cite{mss:02}. 
In both cases, the model was  shown to yield 
predictions in agreement with the data \cite{f2d3dat, DVCSdat}
with no adjustable parameters. The model can also be extended   
to diffractive vector meson production:
\begin{equation}
\gamma^*(q) + p(p) \rightarrow V(q^{\prime}) + p(p^{\prime}) 
\hspace{1cm} V = \rho, \phi~\mbox{or}~J/\Psi \;,
\label{vmp}
\end{equation}
where the squared centre of mass energy
$s = W^2 = (p + q)^{2} \gg Q^2, M_V^2$. In these processes 
the choice of vector meson, as well as of 
different photon virtualities, allows one to explore contributions from dipoles
of different transverse sizes. 
The process also has the advantage that there is a wide range of available
data. It ought also to provide important information on the poorly
known light-cone wavefunctions of the the vector mesons.

The aim of this paper is 
to confront the predictions of the dipole model with the HERA data on $\rho$
and $\phi$ electroproduction. The predictions for $J/\Psi$ production are 
best considered in conjunction with an analysis of open charm production,
and will be discussed  elsewhere. We shall focus on the FKS model  
\cite{fks1}, but we also compare with the predictions of two other models: the 
Golec-Biernat-W\"usthoff (GW) saturation model \cite{gw:99a,gw:99b};
and the recent ``Colour Glass Condensate'' (CGC) model of Iancu, Itakura
and Munier \cite{iim03}.  For
the meson light-cone wavefunction, we shall consider three
different ans\"atze: the Dosch, Gousset, Kulzinger and Pirner 
(DGKP) \cite{dgkp:97} model; the
Nemchik, Nikolaev, Predazzi and Zakharov (NNPZ) \cite{nnpz:97} model;
and a simple ``boosted Gaussian'' wavefunction, which can be considered
as a special case of the latter.

The paper is laid out as follows. In the first two sections we  
summarise the dipole models used and  discuss the 
forms chosen for the vector meson wavefunctions.  We then compare 
their predictions  with experiment  before drawing our 
conclusions.

\section{The colour dipole model}

 In the colour dipole model \cite{dipole}, the eigenstates of the scattering 
(diffraction) operator are ``colour dipoles'', i.e. quark-antiquark pairs
 of transverse size $r$ in which the quark carries a fraction $z$
 of the photon's light-cone momentum\footnote{We work in light-cone 
 coordinates $x^{\mu}=(x^{+},x^{-},\bm{x})$ in the convention where
$x^{\pm}= x^{0}\pm x^{3}$. Here $z=k^{+}/q^{+}$ where the
momentum $k$ of the quark 
is $(k^{+},k^{-},\bm{k})$.}.  In the proton's rest frame, the formation of 
the dipole occurs on a timescale far longer than that of its interaction 
with the target proton. Because of this, the forward imaginary amplitude
for singly diffractive photoprocesses $\gamma p \rightarrow X p$ is assumed
to factorise into a product of
light-cone wavefunctions associated with the initial and final state 
particles $\gamma$ and $X$ and a universal dipole cross-section 
$\hat{\sigma}(s,r)$, which 
contains all the dynamics of the interaction of the $q\bar{q}$ dipole with the target proton. In particular for reaction (\ref{vmp}) one obtains 
\begin{equation}
\Im \mbox{m} \mathcal{A}(s,t=0) = s \sum_{h, \bar{h}}
\int d^2 {\mathbf r} dz \Psi^\gamma_{h, \bar{h}}(r,z)
\hat{\sigma}(s,r) \Psi^{V*}_{h, \bar{h}}(r,z) \; ,
\label{amplitude}
\end{equation}
where $\Psi^{\gamma}_{h, \bar{h}}(r,z)$ and $\Psi^{V}_{h, \bar{h}}(r,z)$ 
are the light-cone wavefunctions  of the photon and vector meson respectively.
The quark and antiquark helicities are labelled by $h$ and $\bar{h}$ and we
have suppressed reference to the meson and photon helicities. 
The dipole cross-section is usually assumed to be flavour independent\footnote
{For the GW model, this is only strictly so at large $Q^2$ since
some flavour dependence enters indirectly at small $Q^2$ through the
definitions of $x_{\mathrm{mod}}$ (see below).} and, as
implied by our notation,  
``geometric'', i.e. for a given $s$, it is assumed to depend on the 
transverse
dipole size, but not the light-cone momentum fraction $z$. 
The light-cone wavefunctions do depend on the quark flavour, via their charges and
masses. Finally, the 
corresponding real part of the amplitude (\ref{amplitude}) is either neglected
or, as here, is estimated using analyticity.

While the photon light-cone wavefunction can be calculated within perturbation
theory, at least for small dipole sizes, the vector meson light-cone 
wavefunctions are not reliably  known, and must be obtained from models.
This will be discussed in the following section. The rest of this section
is devoted to the dipole cross-section, for which we shall consider 
three  different 
models\footnote{For a more general review of 
phenomenological dipole
models, see for example \cite{amirim}.}. Since full details are given in
the original papers, our treatment will be brief. 
  
\subsection{The FKS model}
The FKS model \cite{fks1,fks2,mss:02} is a two-component model
\begin{equation}
\label{gammatot}
{\hat \sigma} (s, r)  =  {\hat \sigma}_{\mathrm{soft}} (s, r) +
{\hat \sigma}_{\mathrm{hard}}(s, r) \, ,  
\end{equation}
\noindent in which each term has  a Regge type energy dependence on the 
dimensionless energy variable $r^2 s$: 
\begin{equation}
{\hat \sigma}_{\mathrm{soft}} (s,r)= a_{0}^{S} 
\left(1-\frac{1}{1+a_{4}^{S} r^{4}}\right) (r^{2} s)^{\lambda_{S}}
\label{sigmasoft}
\end{equation}
\begin{equation}
{\hat \sigma}_{\mathrm{hard}} (s,r)=(a_{2}^{H} r^{2}+a_{6}^{H} r^{6}) ~\exp (-\nu_{H} r) (r^{2} s)^{\lambda_{H}}
\label{sigmahard}
\end{equation}
This parametric form\footnote{This form, taken from \cite{mss:02}, 
is actually a
simplified form of that used in the \cite{fks1,fks2}, but gives almost
identical results.} is chosen so that the hard term dominates at small $r$ 
and goes to zero like $r^2$ as $r \rightarrow 0$ in accordance with ideas of 
colour transparency,
while the soft term dominates at larger $r \approx 1$ fm, with a 
hadron-like soft pomeron behaviour. In addition, to allow for possible
confinement effects  in the  photon wavefunction at large $r$, FKS 
modified the  perturbative wavefunctions $\psi_{T,L}^0(r,z)$ 
by multiplying them by an adjustable Gaussian enhancement factor:
\begin{equation}
  |\psi_{T,L}(r,z)|^{2} = |\psi_{T,L}^0(r,z)|^{2} \, f(r)
\label{peak1}
\end{equation}
where
\begin{equation}
  f(r) = \frac{1 + B \exp(- c^{2} (r - R)^{2})}{1 + B \exp(- c^{2} R^{2})} \; .
\label{peak2}
\end{equation}
This behaviour is qualitatively suggested by an analysis \cite{FGS} of the
scattering eigenstates in a 
generalised vector dominance model \cite{GVD1} which 
provides a good description of the soft Pomeron 
contribution to the nucleon structure function $F_2$ on both protons and
nuclei \cite{GVD2}\footnote{For a more recent discussion of the relation
between GVD models and the dipole approach, see \cite{GVD3}.}.
The  free parameters in both the dipole cross-section and the 
photon wavefunction were then determined  by a 
fit  to structure function and real photoabsorption data. The resulting values 
are given in Table 1. Having been obtained in this way, 
they were then used to predict successfully the
cross-sections for other processes 
which depend solely on the dipole cross-section
and the photon wavefunction, namely diffractive deep inelastic scattering 
(DDIS) \cite{fks2} and Deeply Virtual Compton Scattering (DVCS) \cite{mss:02}. 

\begin{figure}[htbp]
  \begin{center}
   \includegraphics[width=10cm]{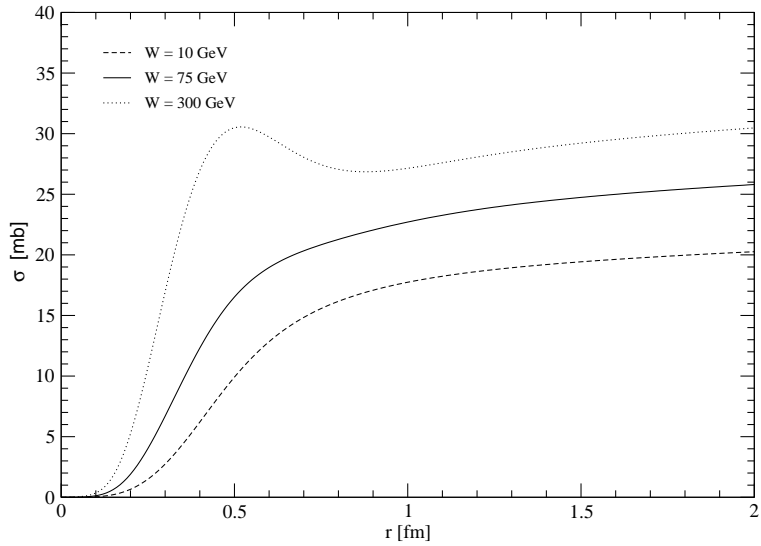}
    \caption{The FKS dipole cross section at $W = 10, 75, 300$~GeV.} 
    \label{fig:fksdipole}
  \end{center}
\end{figure}

The resulting dipole cross-section is shown   
Fig. \ref{fig:fksdipole}. As can be seen, as $s$ increases
the dipole cross-section grows most rapidly for small $r$, where the hard term
dominates, eventually exceeding the typically
hadronic cross-section found for dipoles of large $r \approx 1$ fm. 
This rise could well be tamed by unitarity or saturation effects 
\cite{levin}. However, the authors have argued \cite{fks3}
that such saturation effects are unlikely to be important until
the top of the HERA range and beyond, and they are not included in the FKS
model in its present form. 
  
\begin{table}[htbp]
  \begin{center}
{\bf Table 1}
\[
    \begin{array}{c|c|c|c} 
\hline
\lambda_{S}  & 0.06 \pm 0.01 & \lambda_{H} & 0.44 \pm 0.01 \\
            &              &              &    \\ 
 a_{0}^{S}    & 30.0 \mbox{ (fixed)} & a_{2}^{H}  & 0.072 \pm 0.010  \\
             &              &              &    \\
 a_{4}^{S}    & 0.027 \pm 0.007   &  a_{6}^{H}   & 1.89  \pm 0.03 \\
             &              &              &    \\
             &              &  \nu_{H}  &  3.27 \pm  0.01 \\
             &              &              &    \\
 B            & 7.05 \pm 0.08  &  c^{2}  & 0.20 \mbox{ (fixed)} \\
            &              &              &    \\ 
R            & 6.84 \pm 0.02  &          &  \\
 m_{u,d,s}^2 & 0.08  \mbox{ (fixed)} &  m_{c}^2  
                  & 1.4  \mbox{ (fixed)} \\
\hline  
    \end{array}
\]

    \caption{Parameters for the FKS model \cite{mss:02} 
in appropriate GeV based units.}
    \label{tab:fks-parameters}
  \end{center}
\end{table}

\begin{figure}[htbp]
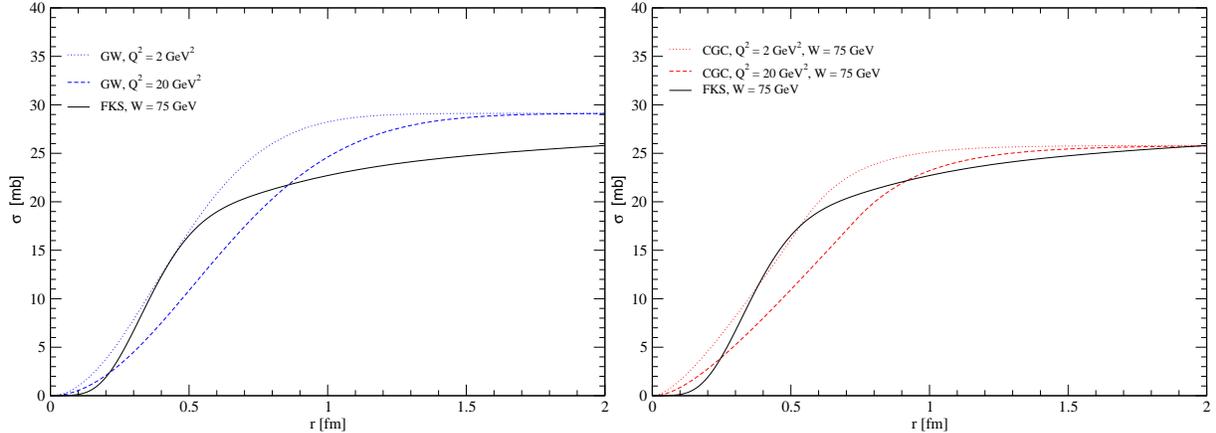

\begin{center}
\includegraphics*[width=8cm]{gw_dipole.eps}\includegraphics*[width=8cm]{cgc_dipole.eps}
\caption{The GW dipole cross-section (left) and CGC dipole cross-section (right) at $W = 75$ GeV for $Q^{2}=
2~\mbox{GeV}^{2}$ 
and $Q^{2}= 20~\mbox{GeV}^{2}$. The $Q^2$-independent
FKS dipole cross-section at the same energy is shown for 
comparison.} 
\label{gw_cgc_dipole.eps}
\end{center}
\end{figure}

\subsection{The GW  model}

This well-known model \cite{gw:99a,gw:99b} combines the approximate behaviour 
$\hat{\sigma} \rightarrow  r^2 f(x)$ at small $r$ together with
a phenomenological saturation effect by adopting the attractively simple 
parametric form:
\begin{equation}
\hat{\sigma} = \sigma_{0}\left(1-\exp \left[\frac{-r^{2}Q_{0}^{2}}
{4(x_{\mathrm{mod}}/x_{0})^{\lambda}}\right]\right) \;.
\label{gbw-dipole}  
\end{equation}
Here $x_{\mathrm{mod}}$ is a modified Bjorken variable,
\begin{equation}
x_{\mathrm{mod}}= x\left(1 + \frac{4m_{f}^{2}}{Q^{2}}\right) \;,
\end{equation}
where $m_{f}$ is the quark mass and $Q_0 = 1$ GeV. 
The $3$ free parameters $x_{0}$, $\sigma_{0}$
and $\lambda$ were successfully fitted to $F_{2}$ data. The four-flavour
fit which we shall use in this paper yielded $\sigma_{0} = 29.12~\mbox{mb}$,
$\lambda =0.277$ and $x_{0} = 0.41\times 10^{-4}$. The quark masses are chosen
to be $0.14$ GeV for the light quarks and $1.5$ GeV for the charm quark. The
model is also able to describe $F_{2}^{D(3)}$ data \cite{gw:99b}. A recent
refinement of the model takes into account corrections due to 
DGLAP evolution at large $Q^{2}$ \cite{gw:02}, but these are rather small 
corrections and are not included here.
Finally, all these results are obtained with a purely perturbative 
photon wavefunction, which is somewhat enhanced at large $r$-values by the 
use of a lighter quark mass than that used in the FKS model\footnote{See, for example, Fig. 3 of \cite{mss:02}. }.

The resulting behaviour of the dipole cross-section is illustrated  
and compared to that of the FKS model in Figure \ref{gw_cgc_dipole.eps}.

\subsection{The CGC model}
The dipole model of Iancu, Itakura and Munier \cite{iim03} can be thought of
as a development of the Golec-Biernat--W\"usthoff saturation model. 
Though still largely
a phenomenological parameterisation, the authors do claim that it contains
the main features of the ``color glass condensate'' regime, where the 
gluon densities are high and non-linear effects become important.
In particular, they take
\begin{eqnarray}
\hat{\sigma} &=& 2 \pi R^2 {\cal N}_0 \left( \frac{r Q_s}{2} 
\right)^{2\left[\gamma_s + \frac{\ln(2/rQ_s)}{\kappa \lambda \ln(1/x)}\right]} 
\hspace*{1cm} \mathrm{for} \hspace*{1cm} rQ_s \le 2 \nonumber \\
&=& 2 \pi R^2 \{1 - \exp[-a \ln^2(brQ_s)]\} \hspace*{1cm} \mathrm{for} 
\hspace*{1cm} rQ_s > 2~,
\label{cgc-dipole}  
\end{eqnarray} 
where the saturation scale $Q_s \equiv (x_0/x)^{\lambda/2}$ GeV. The
coefficients $a$ and $b$ are uniquely determined by ensuring continuity of
the cross-section and its first derivative at $rQ_s=2$. The leading order
BFKL equation fixes $\gamma_s = 0.63$ and $\kappa = 9.9$. The coefficient
${\cal N}_0$ is strongly correlated to the definition of the saturation
scale and the authors find that the quality of fit to $F_{2}$ data is only weakly dependent
upon its value. For a fixed value of ${\cal N}_0$, there are therefore
three parameters which need to be fixed by a fit to the data, i.e.
$x_0$, $\lambda$ and $R$. 
In this paper, we take $N_0 = 0.7$ and a light quark mass of $m_q = 140$ MeV,
for which the fit values are 
$x_0 = 2.67 \times 10^{-5}$, $\lambda = 0.253$ and $R = 0.641$ fm.

As for the GW dipole, we compare to the FKS dipole at two values of $Q^2$
in Figure \ref{gw_cgc_dipole.eps}.

%[RUBEN: are these the values you used? What values did you get for $a$ and
%$b$? Hopefully you did NOT modify the x variable since the limit of 
%$Q^2 \to 0$ is under control here. Also can you confirm that you converted 
%$R$ from fm to $GeV^{-1}$ when using it.]

\section{Light-cone wavefunctions}

The light-cone wavefunctions $\Psi_{h,\bar{h}}(\bm{r},z)$ in
the mixed representation $(r,z)$ used in the dipole model are obtained from
a two dimensional Fourier transform
\begin{equation}
\Psi_{h,\bar{h}}(r,z)= \int
\frac{d^2 \bm{k}}{(2\pi)^{2}} 
e^{i\bm{k} \cdot \bm{r}} \; \Psi_{h,\bar{h}}(k,z)
\label{fourier}
\end{equation}
of the momentum space light-cone wavefunctions $ \Psi_{h,\bar{h}}(k,z)$,
where the quark and antiquark are in states of definite helicity, $h$ 
and $\bar{h}$ respectively. For transversely ($T$) or longitudinally ($L$) polarised photons, the  momentum space light-cone wave functions themselves
are calculated perturbatively \cite{dgkp:97,bl:80} (per fermion of charge
$e e_f$): 
\begin{equation}
\Psi^{\gamma (\lambda)}_{h,\bar{h}}(k,z) = \sqrt{\frac{N_c}{4 \pi}}
\frac{\bar{u}_{h}(\bm{k})}{\sqrt{z}}(ee_{f}\gamma.\varepsilon_{\gamma}^{\lambda})\frac{v_{\bar{h}}(-\bm{k})}{\sqrt{1-z}} \Phi^{\gamma}(k,z) \;. 
\end{equation}
Hereafter, $\lambda$ denotes the polarization state $L$ or $T$.  $\varepsilon^{\lambda}$ are the polarisation vectors and the ``scalar'' 
part\footnote{This would indeed be the photon light-cone 
wavefunction in a toy model of  scalar quarks and photons.} of the photon 
light-cone wavefunction, $\Phi^{\gamma}$, is given by
\begin{equation}
\Phi^{\gamma}(k,z) = \frac{z(1-z)}{z(1-z)Q^2 + k^{2} +  m_f^2}
\;.
\end{equation}
%The quark and antiquark are in states of definite helicity, $h$ and $\bar{h}$ 
%respectively,  $Q^2$  is the photon's virtuality and $m_{f}$ is the 
%appropriate  quark mass. 

% Helicities defined after Eq (14). m_f and Q^2 used  earlier

For the vector mesons, the simplest approach is to assume the same vector 
current as in the photon case, with an additional (unknown) 
vertex factor $\Gamma_{\lambda}(k,z)$: 
\begin{equation}
\Psi^{V(\lambda)}_{h,\bar{h}}(k,z) = \sqrt{\frac{N_c}{4 \pi}} 
\frac{\bar{u}_{h}(\bm{k})}{\sqrt{z}}(\gamma.\varepsilon^{\lambda}_{V})
\frac{v_{\bar{h}}(-\bm{k})}{\sqrt{1-z}}\Phi^{V}_{\lambda}(k,z) 
\label{vwfkz}
\end{equation}
where the scalar part of the meson light-cone wavefunction is given by
\begin{equation}
\Phi^{V}_{\lambda}(k,z) = \frac{z(1-z) \Gamma_{\lambda}(k,z)}{-z(1-z)M_{V}^2 + 
k^{2} + m_f^2} \;.
\label{scalar-meson}
\end{equation}
Different models are defined by specifying these scalar wavefunctions. In
practice, it is common to choose the same functional form for 
$\Phi^{V}_T$ and $\Phi^{V}_L$; perhaps allowing the numerical parameters to 
differ.

It is instructive to consider the longitudinal  wavefunctions more explicitly. Using the 
polarisation vectors
\begin{equation}
\varepsilon_{\gamma}^{L}=\left(\frac{q^{+}}{Q},\frac{Q}{q^{+}},\bm{0}\right)\hspace{0.5cm};\hspace{0.5cm}\varepsilon_{V}^{L}=\left(\frac{v^{+}}{M_{V}},-\frac{M_{V}}{v^{+}},\bm{0}\right)
\label{polarisation-vectors}
\end{equation}
and  the rules of light-cone perturbation theory given in \cite{bl:80}, it follows 
that the longitudinal photon light-cone wavefunction is 
\begin{equation}
\Psi_{h,\bar{h}}^{\gamma,L}(k,z) = \sqrt{\frac{N_c}{4 \pi}}
\delta_{h,-\bar{h}}e e_{f}\left(\frac{2z(1-z)Q}{k^{2} + m_{f}^{2} + 
z(1-z)Q^{2}} - \frac{1}{Q}\right) \;.
\label{photonwf-L}
\end{equation}
and that of the vector meson is
\begin{equation}
\Psi^{V,L}_{h,\bar{h}}(k,z) = \sqrt{\frac{N_c}{4 \pi}}
\delta_{h,-\bar{h}}\left(\frac{z(1-z)2M_{V}\Gamma(k, z)}{k^{2} + m_{f}^{2} - z(1-z)M_{V}^{2}} + \frac{\Gamma(k,z)}{M_{V}}\right) \;.
\label{mesonwf-L}
\end{equation}
On substituting (\ref{photonwf-L}) in (\ref{fourier})
the second term of (\ref{photonwf-L}) leads to a dipole of vanishing size,
which does not contribute to the cross-section. 
This is in accord with gauge invariance.
The same argument cannot be used to 
justify the omission of the second term in the meson wavefunction  
 (\ref{mesonwf-L}),  since the latter has a $k$ dependence. 
In practice, this term is omitted in the DGKP model \cite{dgkp:97},
but retained in  the NNPZ model \cite{nnpz:97}. A discussion of the
gauge invariance issues surrounding this point can be found in 
\cite{hl:98}.

Before discussing these models more fully, we give
the explicit forms for the photon wavefunctions in $r$-space.

The normalised photon light-cone wavefunctions are 
\cite{dgkp:97}:
\begin{equation}
\Psi^{L}_{h,\bar{h}}(r,z) = \sqrt{\frac{N_{c}}{4\pi}}\delta_{h,-\bar{h}}e e_{f}2 z(1-z) Q \frac{K_{0}(\epsilon r)}{2\pi}\;,
\label{photonwfL}
\end{equation}
and 
\begin{equation}
\Psi^{T(\gamma=\pm)}_{h,\bar{h}}(r,z) = \pm \sqrt{\frac{N_{c}}{2\pi}} ee_{f}
 \big[i e^{ \pm i\theta_{r}} (z \delta_{h\pm,\bar{h}\mp} - 
(1-z) \delta_{h\mp,\bar{h}\pm}) \partial_{r}  
+  m_{f} \delta_{h\pm,\bar{h}\pm} \big]\frac{K_{0}(\epsilon r)}{2\pi}\;,
\label{photonwfT}
\end{equation}
where 
\begin{equation}
\epsilon^{2} = z(1-z)Q^{2} + m_{f}^{2}~.
\end{equation}
Since the modified Bessel function $K_{0}(x)$ decreases exponentially 
at large $x$, 
these equations imply  that at high $Q^{2}$, the wavefunctions are suppressed 
for large $r$ unless $z$ is close to its
end-points values $0$ or $1$. As can be seen from equations
(\ref{photonwfL})--(\ref{photonwfT}), these end-points 
 are suppressed for the longitudinal but not the transverse case. This is
the origin of the statement that transverse meson production is more
inherently non-perturbative than for longitudinal meson production. 

For small $r$, the perturbative expressions given above are reliable. 
For large $r$-values, however,  confinement corrections are likely to
modify the perturbation theory result. These larger $r$-values contribute
significantly at   low
 $Q^{2}$, where the wavefunctions are sensitive to the non-zero quark masses
  $m_{f}$,  which  prevent the
 modified Bessel function from diverging in the photoproduction limit. 
 For these reasons, the photon light-cone wavefunctions at large $r$ are 
clearly model-dependent. 

We now turn back to the meson wavefunctions. These are subject to two 
constraints. The first is the normalisation condition \cite{bl:80, munieretal:01} 
\begin{equation}
1 = 
\sum_{h,\bar{h}}\int \frac{d^{2}\bm{k}}{(2 \pi)^2} \, dz \,
|\Psi^{V(\lambda)}_{h,\bar{h}}(k, z)|^{2} =
\sum_{h,\bar{h}}\int d^{2}\bm{r} \, dz  \,
|\Psi^{V(\lambda)}_{h,\bar{h}}(r, z)|^{2} ~,
\label{normalisation}
\end{equation}
which embodies the assumption that the meson is composed solely of
$q\bar{q}$ pairs. Note that this normalisation is consistent with 
equation (\ref{amplitude}) and differs by a factor $4 \pi$ relative to
the conventional light-cone normalisation. 
 
The second constraint comes from the electronic decay width \cite{dgkp:97,munieretal:01}:
\begin{equation}
e f_V M_V \varepsilon_{\gamma}^{*}.\varepsilon_{V} = 
\sum_{h,\bar{h}}\int \frac{d^{2}\bm{k}}{(2 \pi)^2} \frac{dz}
{z(1-z)} (z(1-z) Q^2 + k^2+m_f^2) \Psi^{V}_{h,\bar{h}}(k,z) 
\Psi^{\gamma *}_{h,\bar{h}}(k,z) \; ,
\label{leptonic}
\end{equation}
where the coupling  $f_V$ of the meson to electromagnetic current can be
determined from the experimentally measured leptonic
width $\Gamma_{V\rightarrow e^{+}e^{-}}$ since
$3 M_{V}\Gamma_{V\rightarrow e^{+}e^{-}} = 4\pi\alpha_{\mathrm{em}}^2 
f_{V}^{2}$.
We shall prefer to implement the constraint directly in terms of the
$r$-space wavefunctions. For our purposes, we can write the meson wavefunctions
in $r$-space as
\begin{equation}
\Psi^{V,L}_{h,\bar{h}}(r,z) = \sqrt{\frac{N_{c}}{4\pi}}
\delta_{h,-\bar{h}}\frac{1}{M_{V}z(1-z)} 
[z(1-z)M^{2}_{V} + \delta \times (m_{f}^{2} -  \nabla_{r}^{2})] \phi_L(r,z) 
\label{nnpz_L}
\end{equation}
where $\nabla_r^2 \equiv \frac{1}{r} \partial_r + \partial^2_r$, and 
\begin{equation}
\Psi^{V,T(\gamma=\pm)}_{h,\bar{h}}(r, z) = \pm
\sqrt{\frac{N_{c}}{4\pi}}\frac{\sqrt{2}}{z(1-z)}[i e^{\pm i\theta_{r}} 
( z \delta_{h\pm,\bar{h}\mp} - (1-z) \delta_{h\mp,\bar{h}\pm}) 
\partial_{r}+ m_{f}\delta_{h\pm,\bar{h}\pm}] \phi_T(r, z).
\label{nnpz_T}
\end{equation}

Note the second term in square brackets  which occurs 
in the longitudinal meson case. This is a direct consequence of keeping the 
second term in (\ref{mesonwf-L}). For the DGKP wavefunctions this term
is absent, i.e. $\delta=0$, whilst NNPZ keep this term, i.e. $\delta=1$. 
In terms of these wavefunctions, (\ref{leptonic}) becomes
(assuming that $r \phi_T(r,z) \to 0$ at $r=0$ and $r=\infty$)
\begin{equation}
f_V M_V = \frac{N_c}{\pi} \hat{e}_f \int_0^1 \frac{dz}{z(1-z)} 
\left.[z(1-z)M^{2}_{V} + \delta \times (m_{f}^{2}-\nabla_{r}^{2})] \phi_L(r,z)
\right|_{r=0} 
\label{longdecay}
\end{equation}
and 
\begin{equation}
f_V M_V = -\frac{N_c}{2\pi} \hat{e}_f \int_0^1 \frac{dz}{[z(1-z)]^2} 
\left. \left[ (z^2+(1-z)^2) \nabla_r^2
 - m_f^2 \right] \phi_T(r,z)  \right|_{r=0} 
\label{trandecay} .
\end{equation}

In computing  $f_\phi$ and all other observables involving the
$\phi$ meson we in all cases take the quark mass to be equal to the light 
quark mass plus 150~MeV.
% slight rewording - too many ``Note that's''

\setcounter{footnote}{0}

\subsection{DGKP meson wavefunction}

In the DGKP approach \cite{dgkp:97}, the $r$ and $z$ dependence of the
wavefunction are assumed to factorise\footnote{Note that the
theoretical analysis of Halperin and Zhitnitsky \cite{halperin-zhitnitsky:97}
shows that such a factorising ansatz must break down at the end-points of $z$. However, since the latter are suppressed in the DGKP wavefunction, 
this has no practical consequence.}. Specifically, the scalar wavefunction 
is given by\footnote{DGKP do not actually include the factor $z(1-z)$ in 
the scalar wavefunction. This is because they define the scalar wavefunction 
to be the r.h.s of (\ref{scalar-meson}) divided by $z(1-z)$.}

\begin{equation}
\phi_{\lambda}^{V}(r,z) = G(r) f_{\lambda}(z) z(1-z) 
\label{dgkp1}
\end{equation}
A Gaussian dependence on $r$ is assumed, that is
\begin{equation}
G(r)= \frac{\pi f_V}{N_c \hat{e}_f M_V}
e^{-\frac{\omega^{2}_{\lambda}r^{2}}{2}} \;,
\end{equation}
and $f_{\lambda}(z)$ is given by the Bauer-Stech-Wirbel model \cite{wsb:85}:
\begin{equation}
f_{\lambda}(z)= \mathcal{N}_\lambda 
\sqrt{z(1-z)}e^{\frac{-M_{V}^{2}(z-1/2)^{2}}{2\omega^{2}_{\lambda}}} \;.
\end{equation}
 
Setting $\delta = 0$ (recall that this is 
equivalent to neglecting the second term in (\ref{mesonwf-L})) in 
(\ref{nnpz_L}) results in
\begin{equation}
\Psi_{h,\bar{h}}^{V,L}(r, z) =  
z(1-z) \delta_{h,-\bar{h}}\frac{\sqrt{\pi} f_V}
{2\sqrt{N_{c}}\hat{e}_{f}}f_{L}(z)e^{-\omega_{L}^{2}r^{2}/2} \;,
\label{dgkp_L}
\end{equation}
for the DGKP longitudinal meson light-cone wavefunction. $\hat{e}_f$ is
the effective charge arising from the sum over quark flavours in the meson:
$\hat{e}_f = 1/\sqrt{2}, 1/3$ and $2/3$ for the $\rho,~\phi$ and $J/\Psi$
respectively.
Similarly, the DGKP transverse meson light-cone wavefunctions can be written as 
\begin{eqnarray}
\Psi_{h,\bar{h}}^{V,T(\gamma = \pm)}(r, z) &=& \pm
\left(\frac{i\omega_T^{2}r e^{\pm i\theta_{r}}}{M_{V}}
[z\delta_{h\pm,\bar{h}\mp} - 
(1-z)\delta_{h\mp,\bar{h}\pm}] + \frac{m_{f}}{M_{V}}\delta_{h\pm,\bar{h}\pm}
\right) \nonumber \\ & & \hspace*{5cm} \times 
\frac{\sqrt{\pi} f_V}{\sqrt{2 N_{c}}
\hat{e}_{f}}f_{T}(z)e^{-\omega_{T}^{2}r^{2}/2} \;.
\label{dgkp_T}
\end{eqnarray}

The normalisation condition (\ref{normalisation}) on the DGKP wavefunction 
leads to the relations,
\begin{equation}
\omega_{\lambda} = \frac{ \pi f_{V}}{\sqrt{2 N_c} \hat{e}_{f}}
\sqrt{I_{\lambda}}\;,
\label{dgkp-normalisation}
 \end{equation}
with
\begin{equation}
I_{L} = \int_{0}^{1} dz z^{2}(1-z)^{2} f_{L}^{2}(z) \;,
\end{equation}
and
\begin{equation}
I_{T} =\int_{0}^{1} dz \frac{[z^{2} + (1-z)^{2}]\omega_{T}^{2} + m_{f}^{2}}{M_{V}^{2}} f_{T}^{2}(z) \;,
\end{equation}

The leptonic decay width constraints (\ref{longdecay}) and (\ref{trandecay}) 
on the DGKP wavefunction yield
\begin{equation}
1=\int_{0}^{1} dz \; z(1-z)f_{L}(z)= \int_{0}^{1} dz \frac{2[z^{2} + (1-z)^{2}]\omega_{T}^{2} + m_{f}^{2}}{2M_{V}^{2}z(1-z)}f_{T}(z) \;.
\label{dgkp-leptonic}
\end{equation}
The values of $\omega_{\lambda}$ and $\mathcal{N}_{\lambda}$ are found by
solving (\ref{dgkp-normalisation}) and (\ref{dgkp-leptonic}) simultaneously,
and  are given in Table \ref{tab:dgkp-parameters}.
The values of the decay constants used are the central 
experimental values \cite{pdg}: $f_{\rho} =0.153 \pm 0.004$ GeV, and 
$f_{\phi} =0.079 \pm 0.001$ GeV.
% and $f_{J/\Psi} = 0.273 \pm 0.005$.

\begin{table}[htbp]
  \begin{center}
{\bf Table 2}
\[
    \begin{array}{c|c|c|c|l} 
\hline
\mbox{DGKP} & \omega_{L}  &  \omega_{T}   &  \mathcal{N}_{L}& ~~~~\mathcal{N}_{T}\\
\hline   
\rho&0.331 & 0.206, 0.218 & 15.091 & 5.573, 8.682 \\
%\phi &0.368 & 0.263,0.270 & 15.70 & 8.132, 11.641 \\
\phi&0.368&0.244,0.262&15.703&5.689,8.000\\
%J/\Psi &0.682& 0.590,0.553 & 19.03& 9.989, 8.156\\
% J/\Psi &0.688& 0.593,0.556 & 18.941& 9.900, 8.087\\
\hline       
\end{array}
\]
 \caption{Parameters and normalisations of the  DGKP light-cone wavefunctions 
in appropriate GeV based units. For the transverse case, the first and second 
values are for the FKS and GW quark masses respectively.} 
    \label{tab:dgkp-parameters}
  \end{center}
\end{table}

\begin{figure}[htbp]
\begin{center}
\includegraphics*[width=6.2cm]{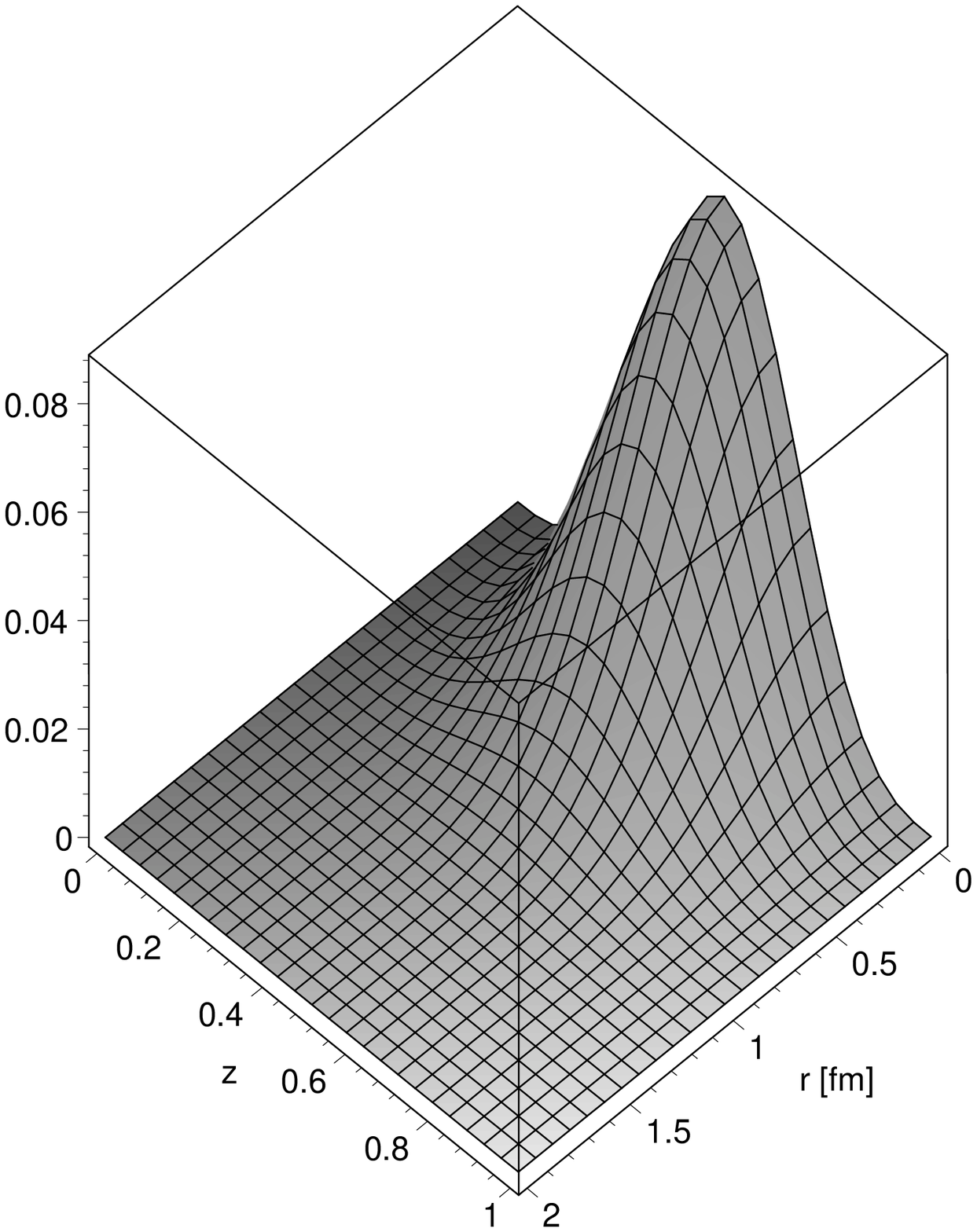} \hspace{2cm} 
\includegraphics*[width=6.2cm]{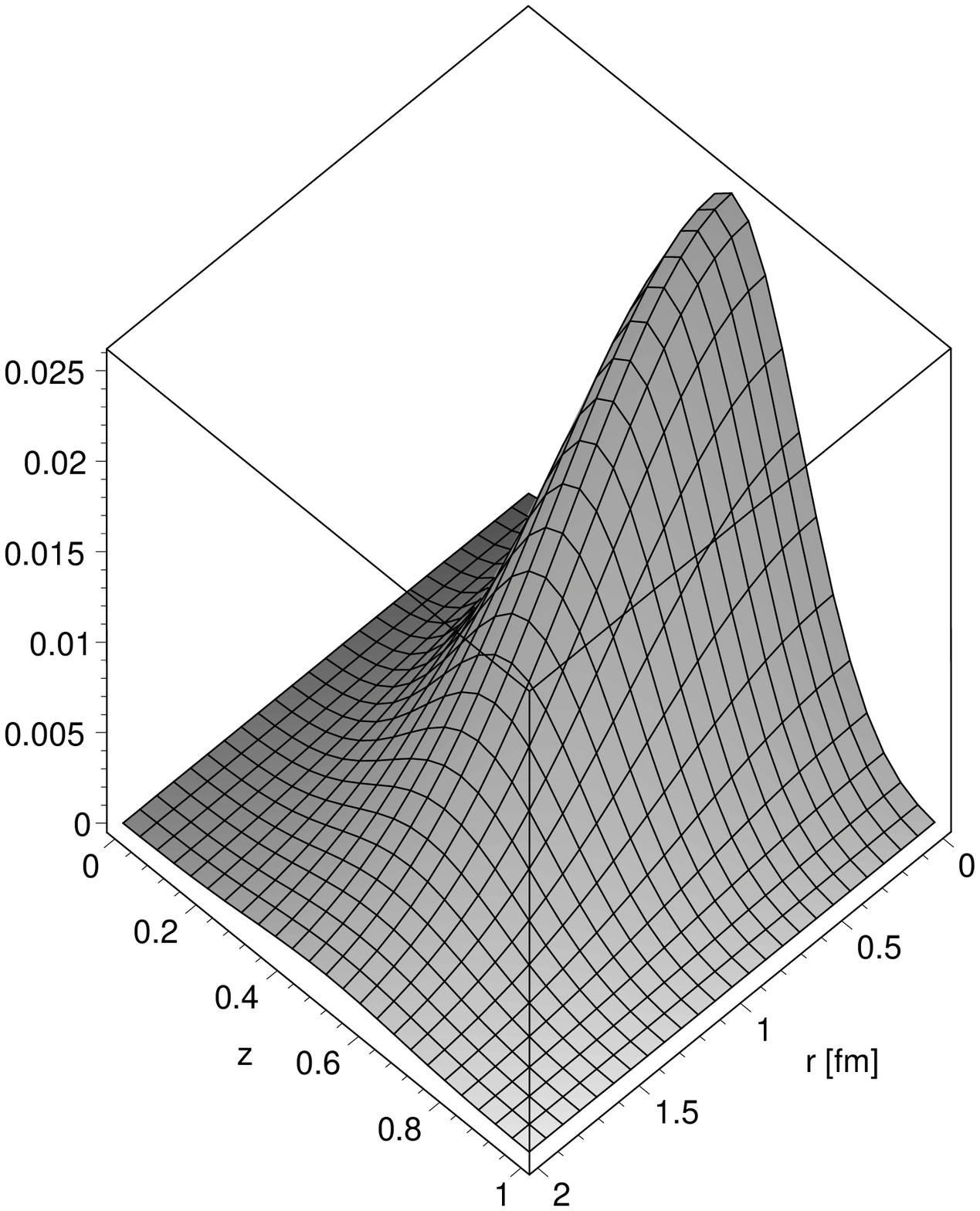}
\caption{The  $\rho$-wavefunctions $|\Psi^{L}|^{2}$ (left)
and $|\Psi^{T}|^{2}$ (right) in the DGKP model with the quark mass
used in the FKS dipole model. Note the different scales for the ordinate.} 
\label{fig:DGKPrho}
\end{center}
\end{figure}

The resulting behaviour of the $\rho$ wavefunctions is shown for the case 
of the FKS quark masses in Figures \ref{fig:DGKPrho}.
% and \ref{fig:DGKPJpsi}. 
As is clear from equations (\ref{dgkp1})--(\ref{dgkp_T}), 
the wavefunctions peak at $z = 0.5$ and $r = 0$, and
go rapidly to zero as $ z \rightarrow 0,1$ and $r \rightarrow \infty$,
so that large dipoles are suppressed. From the figures, we see that
 the transverse wavefunction  has a
broader distribution than the longitudinal wavefunction. 
 The $\phi$ wavefunctions are qualitatively  similar to, but
slightly more sharply peaked than, the $\rho$ wavefunctions.

%\begin{figure}[htbp]
%\begin{center}
%\includegraphics*[width=6.2cm]{jpsi_dgkp_l.eps} \hspace{2cm} 
%\includegraphics*[width=6.2cm]{jpsi_dgkp_t.eps}
%\caption{The  $J/\Psi$-wavefunctions $|\Psi^{L}|^{2}$ (left)
%and $|\Psi^{T}|^{2}$ (right) in the DGKP model with the quark mass
%used in the FKS dipole model. }  
%\label{fig:DGKPJpsi}
%\end{center}
%\end{figure}

\begin{figure}[htbp]
\begin{center}
\includegraphics*[width=6.2cm]{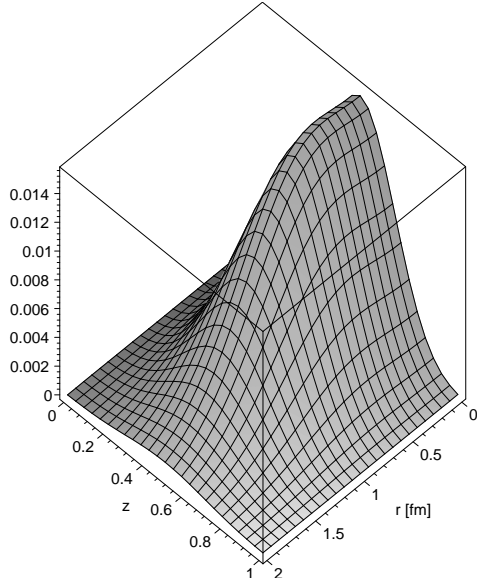}  %\hspace{2cm} 
\caption{The  $\rho$-wavefunction $|\Psi^{T}|^{2}$ in the DGKP model with the 
quark mass used in the GW dipole model. }  
\label{fig:DGKP-GW}
\end{center}
\end{figure}

The GW model uses a much smaller value for the light quark masses than
the FKS model, as we saw in Section 2.3.  We might expect this to have a 
striking effect on the transverse wavefunction of the $\rho$, since 
the transverse wavefunction (\ref{dgkp_T}) vanishes at $r=0$ for zero
quark masses, while the longitudinal  wavefunction (\ref{dgkp_L}) does not. 
The transverse DGKP wavefunction with the light quark mass used in the GW 
dipole model is shown in Figure \ref{fig:DGKP-GW} 
for the $\rho$. As can be seen, the smaller quark mass decreases 
the wavefunction at the origin and shifts the peak to slightly larger $r$.

\subsection{Boosted wavefunctions}

In this approach, the scalar part of the wavefunction is obtained 
by taking a given wavefunction in the meson rest frame. This is then
 ``boosted'' into a light-cone 
wavefunction  using the Brodsky-Huang-Lepage prescription, in   
which the expressions for the off-shellness in the centre-of-mass and
light-cone frames are equated  \cite{bhl:80} (or equivalently, 
the expressions for the invariant mass of the $q \bar{q}$
pair in the centre-of-mass and light-cone frames are equated \cite{dgs:01}). 

The simplest version of this approach assumes a simple Gaussian
wavefunction in the meson rest frame. Alternatively NNPZ 
 \cite{nnpz:97} have supplemented this by adding a hard ``Coulomb''
contribution in the hope of improving the description of the rest frame 
wavefunction at small $r$. We refer to  \cite{nnpz:97} for details of this
procedure. Here we simply state the result, which is that 
the NNPZ meson light-cone wavefunctions are given by
equations (\ref{nnpz_L}, \ref{nnpz_T}) with $\delta = 1$, where  the 
scalar wavefunctions $\phi_\lambda(r,z)$ 
are taken to be a sum of a soft (Gaussian in the rest frame) part and a
hard (Coulomb) part:
\begin{eqnarray}
\phi_\lambda(r,z) &=&
\mathcal{N}_\lambda\ \left(   4z(1-z) \sqrt{2\pi R^{2}}
\exp \left(-\frac{m_f^{2}R^{2}}{8z(1-z)}\right) 
\exp\left(-\frac{2z(1-z)r^{2}}{R^{2}}\right) \right. \nonumber \\
&& \left. \times\exp\left(\frac{m_f^{2}R^{2}}{2}\right) + 
16 C^{4} \frac{a^{3}(r)}{A(r,z) B^{3}(r,z)}r 
K_{1}(A(r,z) r / B(r,z)) \right) .
\label{nnpz-lc-wf} 
\end{eqnarray}
Here
\begin{equation}
A(r,z)= \sqrt{1 + \frac{C^{2}a^{2}(r)m_{f}^{2}}{z(1-z)}-
4C^2 a^{2}(r) m_{f}^{2}} \; ,
\label{Arz}
\end{equation}
\begin{equation}
B(r,z) = \frac{C a(r)}{\sqrt{z(1-z)}} \;,
\label{Brz}
\end{equation}
and 
\begin{equation}
a(r) = \frac{3}{4 m_{f}\alpha_{s}(r)} 
\label{bohr-radius}
\end{equation}
is  a running Bohr radius. The  strong coupling
$\alpha_{s}$ is chosen to run according to the  
prescription \cite{nnpz:94,munieretal:01}:
\begin{equation}
\alpha_{s} = \alpha_0~~\mbox{for}~~ r > r^{s}~~~ \mbox{and} ~~
\alpha_{s}(r) = \frac{4\pi}{\beta_{0}\log
\left(1/(\Lambda_{\mbox{\tiny{QCD}}}^{2}r^{2})\right)}~~\mbox{for}~~
r < r^s
\label{running-alphas}
\end{equation}
where $r^{s} = 0.42$ fm, $\alpha_{0} = 0.8$,
$\Lambda_{\mbox{\tiny{QCD}}}= 0.2~\mbox{GeV}$ and $\beta_{0} = (33 -
2n_{f})/3$. 
Apart from the normalisation constants $ \mathcal{N}_\lambda $ , these
wavefunctions depend on two free parameters which are
independent of the meson helicity: a ``radius'' parameter $R$ 
and a parameter $C$, introduced to control the transition
between the hard Coulomb-like interaction and the soft confining interaction 
in the rest-frame wavefunction. The case of a simple Gaussian wavefunction in
the rest-frame, which we will refer to as a ``boosted Gaussian wavefunction''
can be obtained by simply setting $C = 0$ (we still choose $\delta = 1$ when
considering this wavefunction).
 
At this point we comment on  two issues associated with the
behaviour of the Coulomb part of the scalar wavefunction for small $r$. 
The first is that (\ref{nnpz-lc-wf}) diverges logarithmically as 
$r \rightarrow 0$ at $z = 0.5$. This divergence is however regulated in 
observables. For example, while  the resulting squared wavefunctions
$|\Psi^{V(\lambda)}|^2$ exhibit a narrow, singular  peak at $r =0, z=0.5$,
the quantity $r|\Psi^{V(\lambda)}|^2$ which enters the normalisation condition
(\ref{normalisation}) is zero in this limit. Nevertheless, this singular
behaviour does have a (finite) effect when computing the meson decay constant
which depends upon the behaviour of the wavefunction at $r=0$.
The second issue is that  when 
the scalar wavefunction (\ref{nnpz-lc-wf}) is substituted into equations
(\ref{nnpz_L}) -- (\ref{trandecay}), the derivatives in $r$ 
give rise to inverse power divergences at $r=0$ when acting upon the 
running coupling (\ref{running-alphas}).  
However, these divergences occur solely in terms 
which are strictly higher order in $\alpha_{s}$. Henceforth 
we discard these higher order terms, which is equivalent to omitting all 
derivatives of $\alpha_{s}(r)$ with respect to $r$ when differentiating 
the scalar wavefunction (\ref{nnpz-lc-wf}).
We stress that none of these issues arise when using the boosted Gaussian
wavefunction. 

It remains to determine the various constants. 
NNPZ \cite{nnpz:97} determined both $R$ and  $C$
by using a standard variational procedure for the initial centre-of-mass
wavefunction  using a non-relativistic potential. They then checked that 
the resulting predictions (\ref{leptonic})
were in reasonable accord with the observed leptonic decay 
widths\footnote{Note that a shortcoming of this model is that 
equations (\ref{leptonic})--(\ref{nnpz_T}) 
give slightly different predictions for 
the decay constant for the case of transverse and longitudinal meson 
helicities, because NNPZ use helicity independent
values of $R,C$.}. Here we follow a slightly modified  procedure, since we
 want to be able to easily 
adjust the quark masses to those assumed in the various dipole models.
Specifically, we fixed $C$ at the value chosen by 
NNPZ and vary the value of  $R$  to give approximate agreement
with  the decay width constraints (\ref{leptonic}). 
 In practice, we found it 
adequate to use the same $R$-value for both the FKS and GW mass choices. 
The resulting  values of $R$ and $C$, with the associated values of
the normalisation constants, are shown in Table 
\ref{tab:nnpz-parameters}, and are not very different from the
original parameters of NNPZ. In addition we show the results for 
the boosted Gaussian case ($C = 0$) in Table \ref{tab:gaussian-parameters}. 

\begin{table}[htbp]
  \begin{center}
{\bf Table 3}
\[
    \begin{array}{c|c|c|c|c|c|c} 
\hline
\mbox{NNPZ} & R^{2} &  C & \mathcal{N}_{L} & \mathcal{N}_{T} & f_{V}(L)&
f_V (T) \\
\hline   
\rho &25.0&0.36&0.0123,0.0121&0.0125,0.0137&0.143,0.147&0.157,0.109\\
\phi &18.0&0.53 &0.0122,0.0124&0.0124,0.0131&0.078,0.078&0.087,0.067\\
%J/\Psi&3.6&1.13&0.0150,0.0145&0.0154,0.0146&0.267,0.260&0.249,0.284\\
\hline
\end{array}
\]
\caption{Parameters of the NNPZ light-cone wavefunctions 
in appropriate GeV based units, together with the resulting values 
for the decay constants $f_V$. Where two values are given, 
the first and second values are for the FKS and GW quark masses 
respectively. }
\label{tab:nnpz-parameters}
\end{center}
\end{table}

\begin{table}[htbp]
\begin{center}
{\bf Table 4} 
\[
    \begin{array}{c|c|c|c|c|c} 
\hline
\mbox{Gaussian} & R^{2}& \mathcal{N}_{L} & \mathcal{N}_{T} & f_{V}(L)&
f_V (T)\\
\hline   
\rho &12.3&0.0213,0.0244&0.0221,0.0259&0.153,0.161&0.203,0.192\\
\phi& 10.0&0.0214,0.0243&0.0219,0.0251&0.075,0.079&0.095,0.088\\
%J/\Psi&2.44&0.0388,0.0359&0.0392,0.0361&0.273,0.260&0.277,0.295\\
\hline
\end{array}
\]
\caption{Parameters of the boosted Gaussian wavefunction 
in appropriate GeV based units, together with the resulting values 
for the decay constants $f_V$. Where two values are given, 
the first and second values are for the FKS and GW quark masses 
respectively. }
\label{tab:gaussian-parameters}
\end{center}
\end{table}

\begin{figure}[htbp]
\begin{center}
\includegraphics*[width=6.2cm]{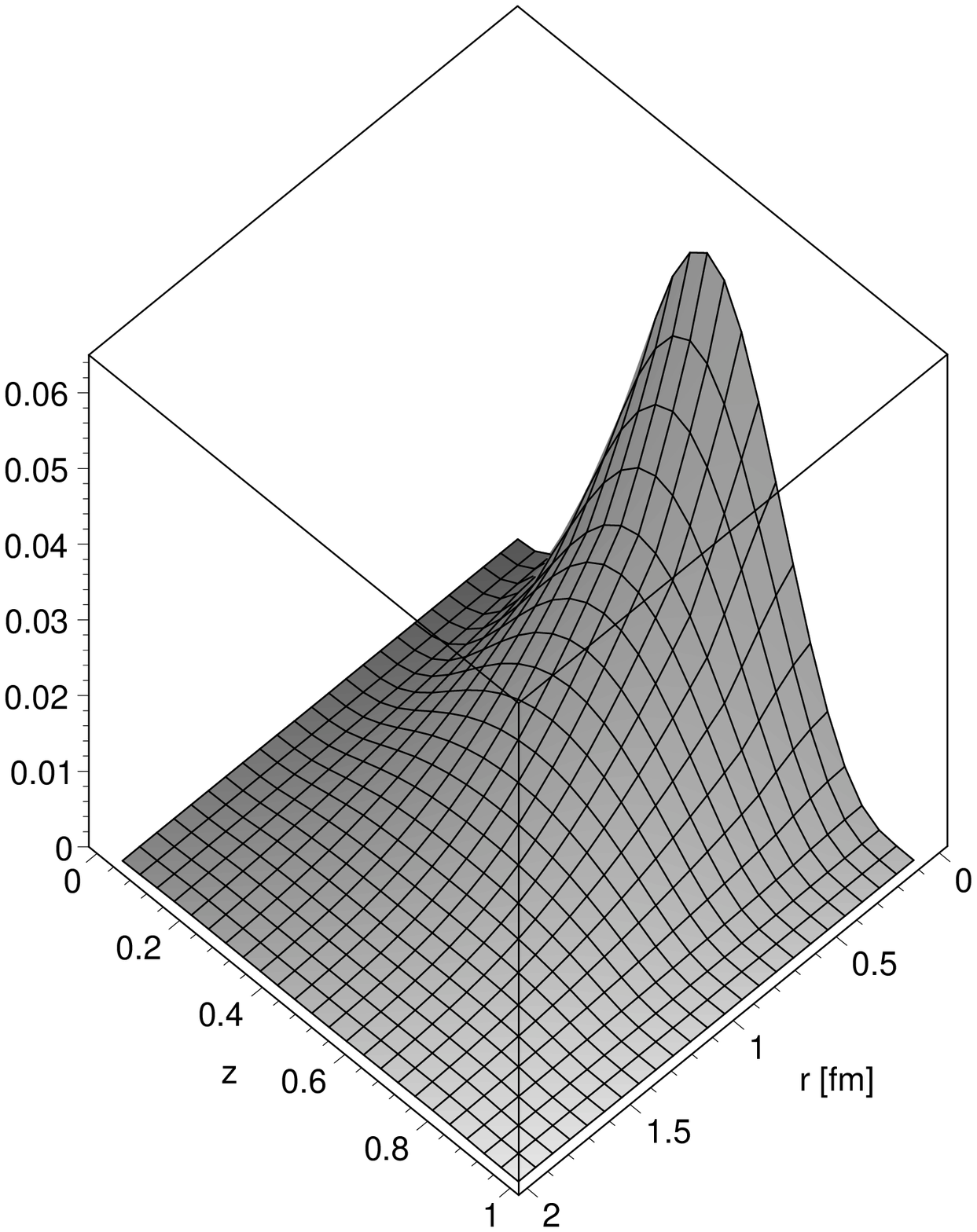} \hspace{2cm} 
\includegraphics*[width=6.2cm]{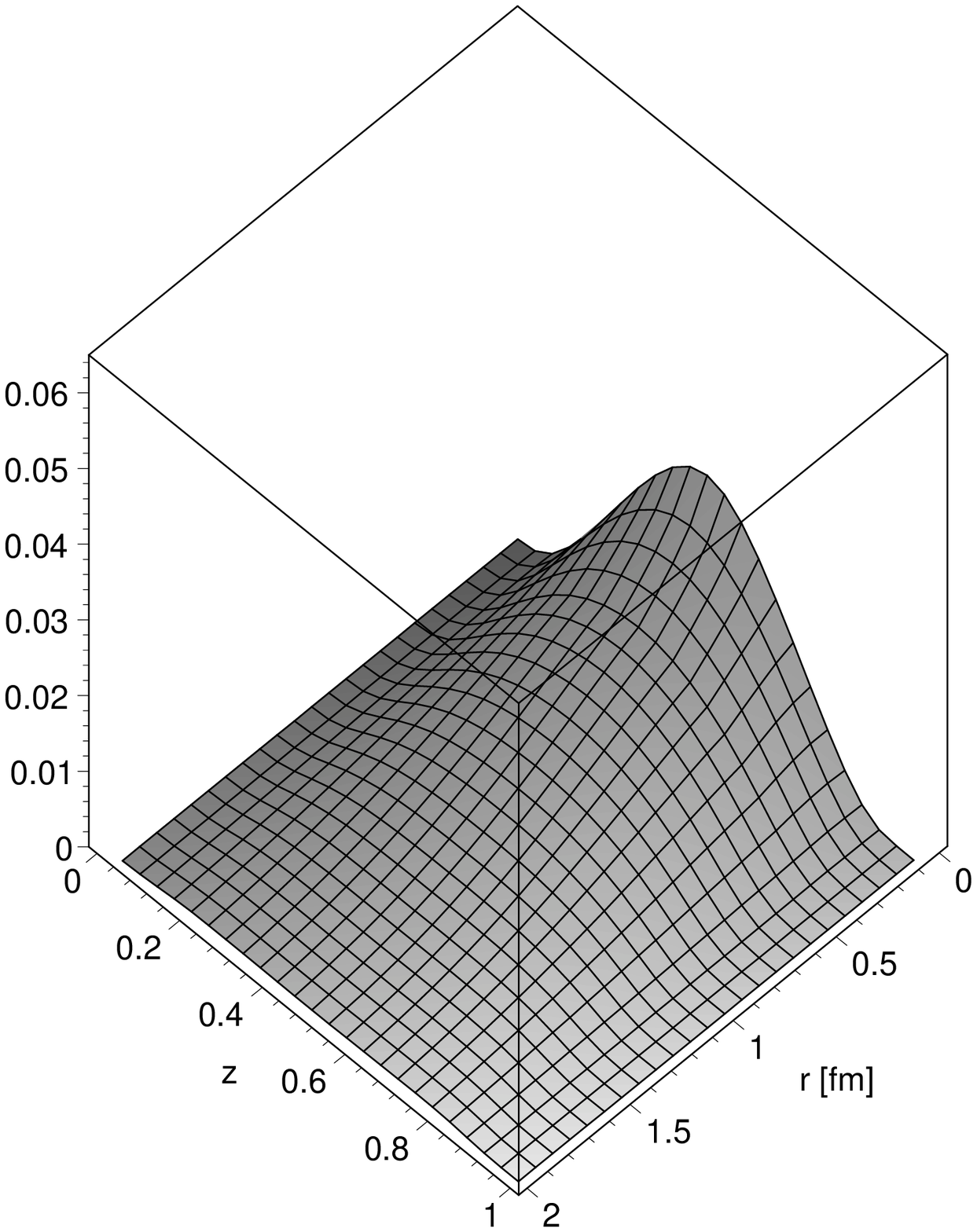}
\caption{The  $\rho$-wavefunctions $|\Psi^{L}|^{2}$ (left)
and $|\Psi^{T}|^{2}$ (right) in the NNPZ model with the quark mass
used in the FKS dipole model.} 
\label{fig:NNPZrho}
\end{center}
\end{figure}

The behaviour of the resulting $\rho$ wavefunctions is shown in Figures
\ref{fig:NNPZrho} with the FKS choice of quark masses. The divergence
at $r=0, z=0.5$ is not visible since we do not plot down to $r=0$.
Like the DGKP wavefunctions, they peak at $z = 0.5$ and $r = 0$, and
go rapidly to zero as $ z \rightarrow 0,1$ and $r \rightarrow \infty$. 
However, on comparing these figures with each other, and with 
Figures \ref{fig:DGKPrho}, 
%and \ref{fig:DGKPJpsi},
two clear differences emerge.

Firstly, the peak in $z$ is less sharp in the boosted Gaussian case
than in the DGKP and NNPZ cases.

Secondly,  the large difference between the 
longitudinal and transverse wavefunctions found in the DGKP case is much less
marked in the NNPZ and boosted Gaussian wavefunctions. 
In both cases, the  peak in the transverse wavefunction is still broader
than that in the longitudinal wavefuction, but it is a small effect in 
comparison with the DGKP case. This presumably reflects the fact that in
the NNPZ and boosted Gaussian wavefunctions, the parameter $R$ has the 
same value for both
helicities, since both wavefunctions are generated from the same 
non-relativistic wavefunction by the Brodsky-Huang-Lepage procedure. 

%In the case of the $J/\Psi$-meson, the peaks are very much narrower in
%NNPZ than in DGKP, with the $z$-dependence approaching the delta-function
%at $z = 0.5$ expected in the extreme non-relativistic case where the
%internal motion of the quarks is neglected completely.

\begin{figure}[htbp]
\begin{center}
\includegraphics*[width=6.2cm]{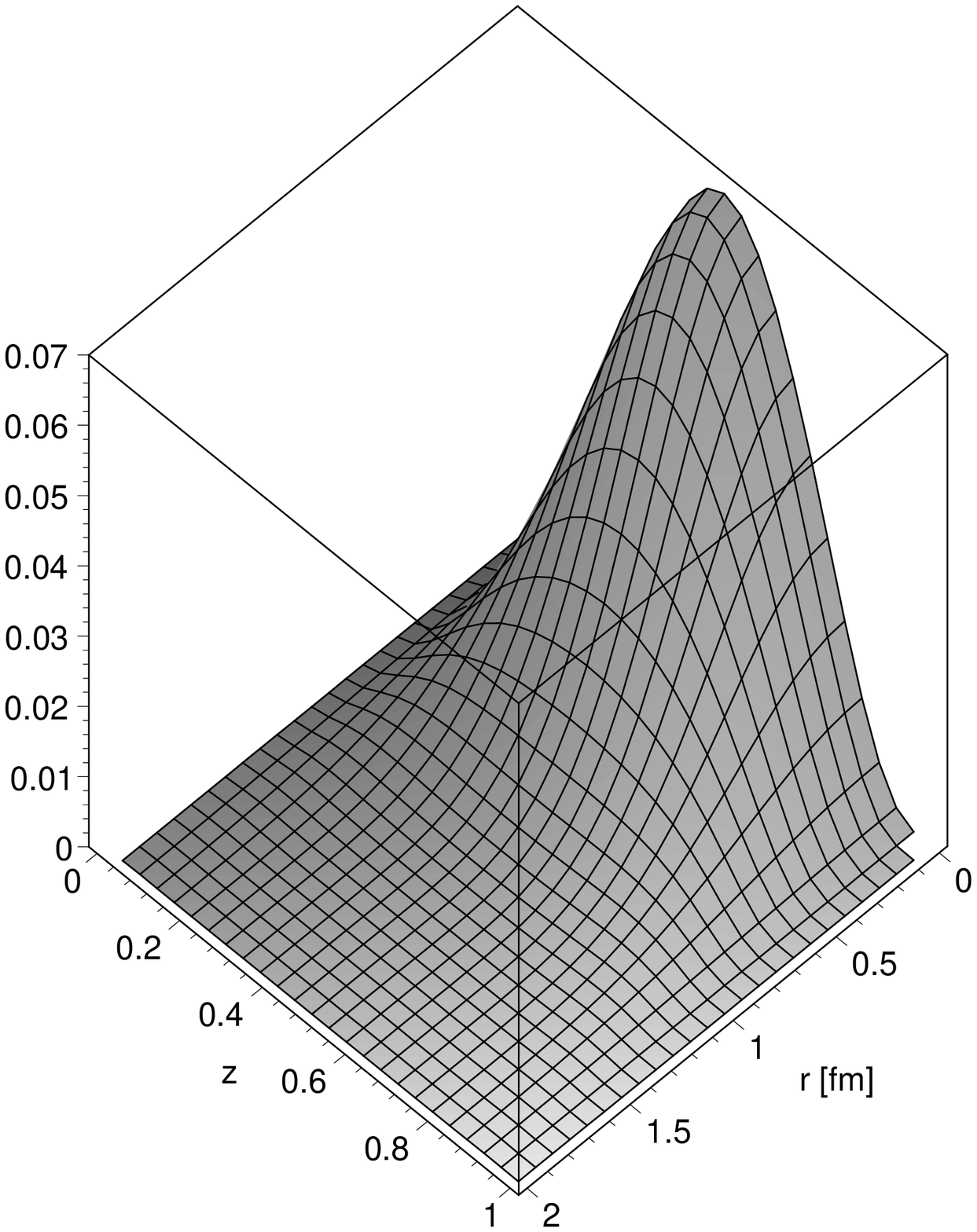} \hspace{2cm} 
\includegraphics*[width=6.2cm]{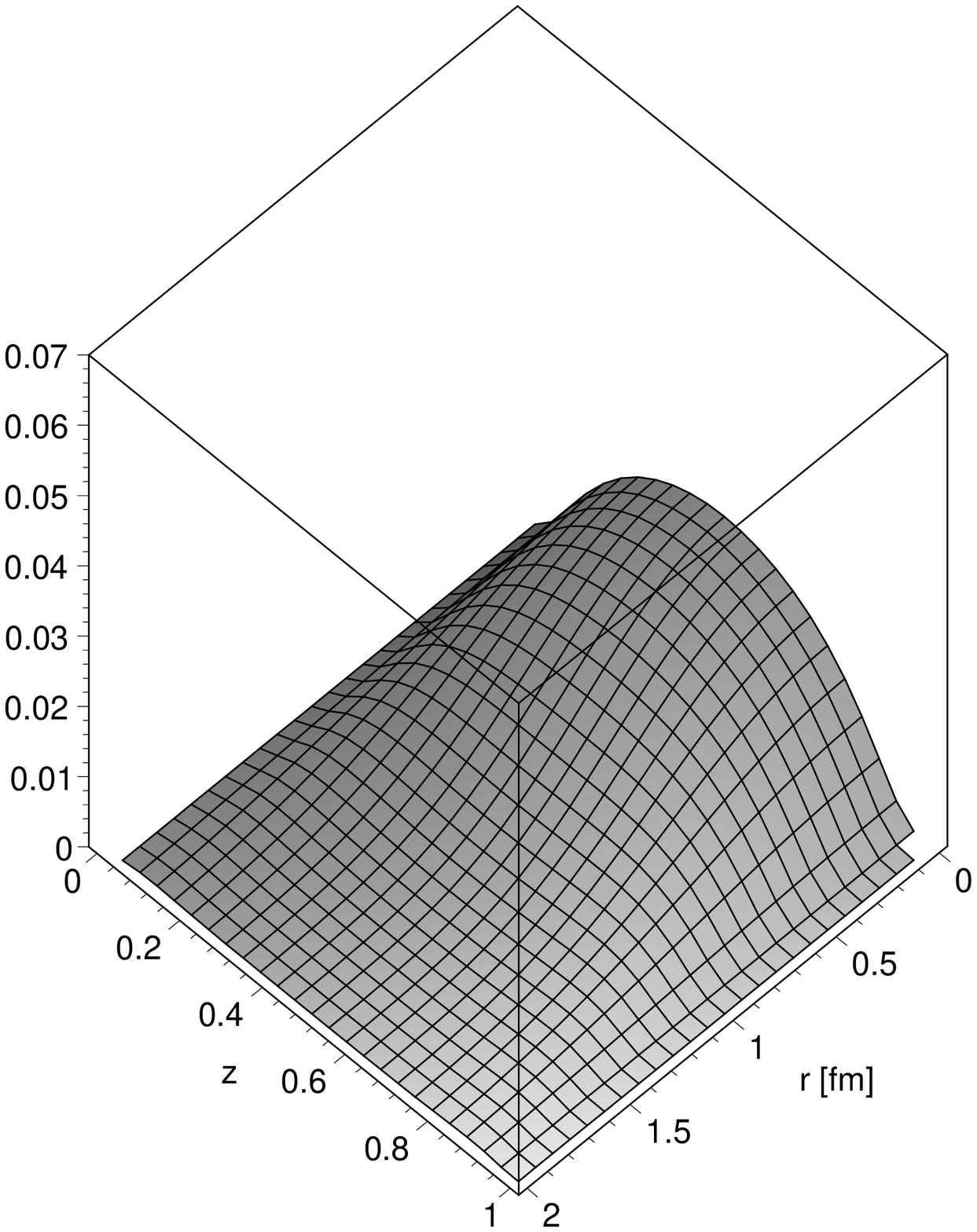}
\caption{The  $\rho$-wavefunctions $|\Psi^{L}|^{2}$ (left)
and $|\Psi^{T}|^{2}$ (right) in the boosted Gaussian model with the quark mass
used in the FKS dipole model.} 
\label{fig:gaussrho}
\end{center}
\end{figure}

Figures \ref{fig:gaussrho} 
%and \ref{fig:gaussjpsi} 
show the wavefunctions in the case of the boosted Gaussian wavefunction.

%\begin{figure}[htbp]
%\begin{center}
%\includegraphics*[width=6.2cm]{jpsi_nnpz_l.eps} \hspace{2cm} 
%\includegraphics*[width=6.2cm]{jpsi_nnpz_t.eps}
%\caption{The  $J/\Psi$-wavefunctions $|\Psi^{L}|^{2}$ (left)
%and $|\Psi^{T}|^{2}$ (right) in the NNPZ model with the quark mass
%used in the FKS and dipole model.}  
%\label{fig:NNPZJpsi}
%\end{center}
%\end{figure}

%\begin{figure}[htbp]
%\begin{center}
%\includegraphics*[width=6.2cm]{g_jpsi_long.eps} \hspace{2cm} 
%\includegraphics*[width=6.2cm]{g_jpsi_trans.eps}
%\caption{The  $J/\Psi$-wavefunctions $|\Psi^{L}|^{2}$ (left)
%and $|\Psi^{T}|^{2}$ (right) in the boosted Gaussian model with the quark mass
%used in the FKS and dipole model.}  
%\label{fig:gaussjpsi}
%\end{center}
%\end{figure}

\section{ Real parts, slope parameters  and cross-sections}
We now have all the ingredients required to calculate the absorptive parts 
 (\ref{amplitude}) of  the forward amplitudes for vector meson
 production. For the case of the NNPZ wavefunctions,  we substitute
the photon wavefunctions\footnote{For the FKS case, we must also include
the effect of the enhancemect factor (\ref{peak2}).} 
 (\ref{photonwfL}), (\ref{photonwfT}) and the vector meson wavefunctions
(\ref{nnpz_L}), (\ref{nnpz_T}) with $\delta = 1$ into (\ref{amplitude}).
Summing over the quark/antiquark helicities  and averaging over the  
transverse polarisation states of the photon, we obtain\footnote{These 
results reduce to equation (8) of \cite{nnpz:97} if, in (\ref{amp-nnpz-L}),
we integrate the second term by parts assuming 
an $r$-independent dipole cross-section.}:
\begin{eqnarray}
\Im \mbox{m} \mathcal{A}^{L}_{\mbox{\tiny{NNPZ}}}&=& s \frac{N_{c}\hat{e}_{f}
\sqrt{4\pi\alpha_{{\mathrm{em}}}}}{(2
\pi)^{2}} \frac{2Q} {M_{V}} \int  d^{2} 
\bm{r} \, \hat{\sigma}(s,r) \,
\int_0^1 dz \nonumber \\ & &
\left\{ [ m_{f}^{2} + z(1-z)M_{V}^{2}]K_{0}(\epsilon r) \phi_{L}(r,z) - 
K_{0}(\epsilon r) \nabla_{r}^{2}) \phi_{L}(r,z) \right\}  
\label{amp-nnpz-L}
\end{eqnarray} 
\begin{eqnarray}
\Im \mbox{m} \mathcal{A}^{T}_{\mbox{\tiny{NNPZ}}} &=& s  
\frac{N_{c}\hat{e}_{f}
\sqrt{4\pi\alpha_{{\mathrm{em}}}}}{(2\pi)^{2}} \int  d^{2}\bm{r}\, 
\hat{\sigma}(s,r) \,
\int_{0}^{1} \frac{dz}{z(1-z)} \nonumber \\ & & \left[
(z^{2} + (1-z)^{2})\partial_{r} K_{0}(\epsilon r) \partial_{r}\phi_{T}(r,z)
 + m_{f}^{2}K_{0}(\epsilon r)\phi_{T}(r,z) \right] \; ,
\label{amp-nnpz-T}
\end{eqnarray}
where $\phi_{L,T}(r,z)$ are given by (\ref{nnpz-lc-wf}).
Similarly, using the DGKP wavefunctions (\ref{dgkp_L}) and  (\ref{dgkp_T}),
we obtain \cite{dgkp:97}
\begin{equation}
\Im \mbox{m} \mathcal{A}^{L}_{\mbox{\tiny{DGKP}}} = 
s \int d^{2}\bm{r}\,\hat{\sigma}(s,r) \int_{0}^{1} dz 
\sqrt{\frac{\alpha_{\mathrm{em}}}{4\pi}} f_{V} z(1-z) f_L(z) 
e^{-\omega_L^{2}r^{2}/2}2z(1-z)QK_{0}(\epsilon r) \;,
\label{amp-dgkp-L}
\end{equation}
\begin{eqnarray}
\Im \mbox{m} \mathcal{A}^{T}_{\mbox{\tiny{DGKP}}} &=& s \int d^{2}{\bm{r}}\,\hat{\sigma}(s,r) \int_{0}^{1} dz \sqrt{\frac{\alpha_{{\mathrm{em}}}}{4\pi}} f_{V} 
f_T(z) e^{-\omega_T^{2}r^{2}/2} \nonumber \\ & &
\left(\frac{\omega_T^{2}\epsilon r}{M_{V}}[z^{2} + (1-z)^{2}] K_{1}(\epsilon r) + \frac{m_{f}^{2}}{M_{V}}K_{0}(\epsilon r) \right) \;.
\label{amp-dgkp-T}
\end{eqnarray}

So far, we have focussed on the imaginary amplitude. Taking into account the real part contribution, the differential cross-section is  given by 
\begin{equation}
\left.\frac{d\sigma^{T,L}}{dt}\right|_{t=0}= \frac{1}{16\pi s^{2}}|\Im \mbox{m}
\mathcal{A}^{T,L}|^{2}(1+\beta^{2}) \;
\label{dcs}
\end{equation}
where $\beta$ is the ratio of real to imaginary parts of the amplitude. It is
most straightforward to reconstruct the real part of the amplitude in the FKS
dipole model, where the dipole cross-section (\ref{gammatot}), and hence the
amplitude  (\ref{amplitude}), is given as the sum of  hard and soft Regge
pole terms. In this case, the real part is given
by

\begin{equation}
\Re \mbox{e} \mathcal{A}_{\mbox{\tiny{FKS}}} = - \Im
 \mbox{m}  \mathcal{A}_{\mbox{\tiny{soft}}} \cot \left( \frac{\pi
 \alpha_{S}}{2} \right) -
 \Im \mbox{m} \mathcal{A}_ {\mbox{\tiny{hard}}}
 \cot \left ( \frac{\pi \alpha_{H}}{2}\right) \;,
\label{ratio}
\end{equation}

where $\alpha_{S,H} = 1 + \lambda_{S,H} = 1.06,1.44$ and
 $\Im \mbox{m}  \mathcal{A}_{\mbox{\tiny{soft}}}$ and $\Im
\mbox{m}  \mathcal{A} _{\mbox{\tiny{hard}}}$ are the contributions from 
 the soft and hard Pomeron pieces of FKS dipole cross-section  
 (\ref{gammatot}) respectively.

\begin{figure}[h]
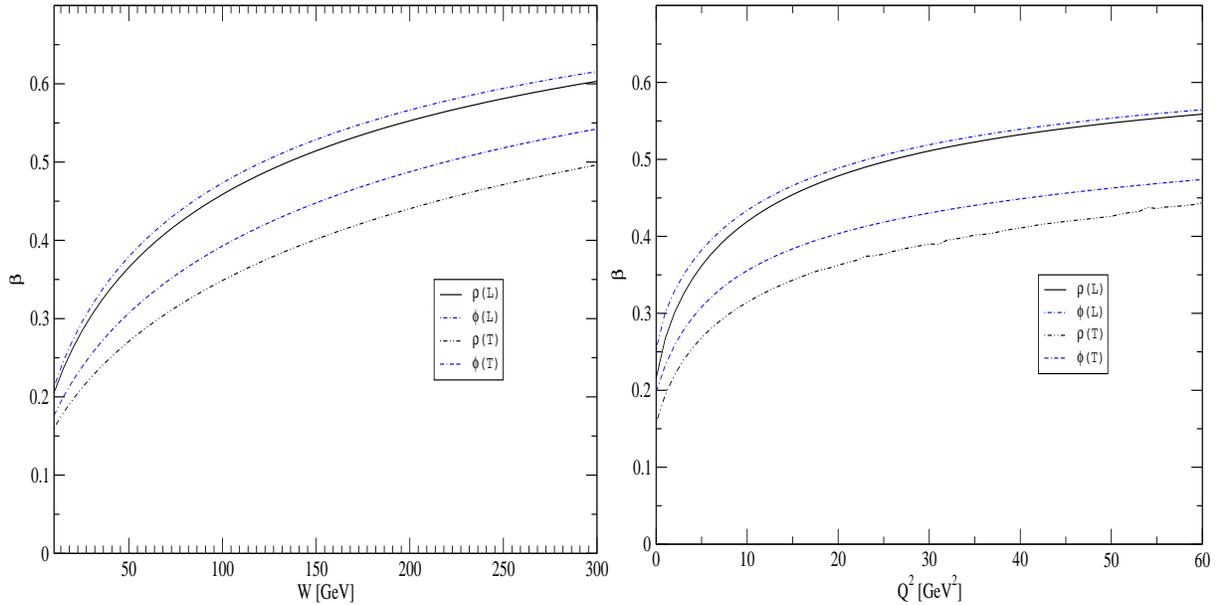

\begin{center}
\includegraphics*[width=8cm,height=8cm]{rp_beta_dgkp_q2_10.eps}\includegraphics*[width=8cm,height=8cm]{rp_beta_dgkp_w_75.eps}
\caption{\textit{Left}: $W$-dependence of the ratio $\beta$ of the real to imaginary part of the forward amplitude in the FKS model, using the DGKP wavefunctions, for $Q^2 = 10$ GeV$^2$. \textit{Right}: $Q^{2}$-dependence  of $\beta$ in the FKS model, using the DGKP wavefunctions, for $W=75$ GeV.}
\label{fig:rp_beta_dgkp_q2}
\end{center}
\end{figure}

It follows from  (\ref{ratio}) that, in the FKS model,
 $\beta$ lies between 0.09 and 0.83, corresponding to pure soft and
 hard pomeron dominance respectively; within this range, the value of
$\beta$ reflects directly the relative importance of the hard pomeron. Other 
things being equal, $\beta$ will therefore increase with increasing energy,
because the hard term in the dipole cross-section increases more rapidly with 
energy than the soft term. It will also increase with $Q^2$, because the hard 
term is dominant for small dipoles, which are increasingly explored as $Q^2$
increases. 
%For the same reason, one would expect it to be larger for the
%$J/\Psi$ than for the $\rho$ and $\phi$, because the $J/\Psi$ wavefunction is 
%peaked more shaply at $r = 0.$  
These features are illustrated in
Figure \ref{fig:rp_beta_dgkp_q2},
which shows the values of $\beta$ obtained as functions of $W$ and $Q^2$
obtained using the DGKP wavefunctions. Here we also see 
that  $\beta$ is larger for the longitudinal than for 
the transverse case, reflecting sharper peaking of the longitudinal vector
meson wavefunctions.
% but that this effect is suppressed for the $J/\Psi$
%where the longitudinal and transverse wavefunctions are less different,
%as discussed in Section 3.2. 
Similar results for the real parts are
obtained using the NNPZ wavefunctions.

From the above figures, we see that while the corrections from the real parts
in the cross-section formulae (\ref{dcs}) are clearly significant in some 
kinematic ranges, they are nowhere dominant. Because of this and because
the ratio $\beta$ is expected to be similar in the different 
models\footnote{This was confirmed explicitly for two distinct dipole models in \cite{mss:02}.}, we shall use the  estimates (\ref{ratio}) of the
 ratio $\beta$ obtained in the FKS model in both dipole models.

\begin{figure}[h]
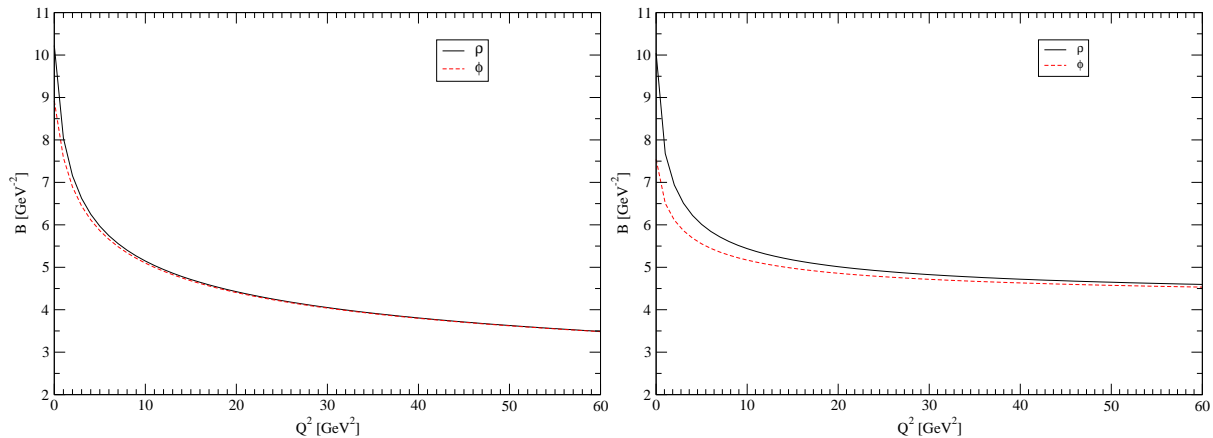

\begin{center}
\includegraphics*[width=8cm]{rp_slope_m.eps}\includegraphics*[width=8cm]{rp_slope_k.eps}
\caption{\textit{Left}: Parametrisation of the slope $B$ by Mellado 
\cite{mellado} for the $\rho$ and $\phi$. 
\textit{Right}: Parametrisation of the slope $B$ by Kreisel 
\cite{kreisel:02} for the $\rho$ and $\phi$.}  
\label{fig:slope}
\end{center}
\end{figure}

Assuming the usual exponential ansatz for the $t$ dependence, the total 
cross-sections are given by
\begin{equation}
\sigma^{T,L}(\gamma^{*}p\rightarrow V p) = \frac{d\sigma^{T,L}/dt|_{t=0}}{B} \;
.
\label{sigmatot}
\end{equation}
Unfortunately, the values 
of the slope parameter $B$ are not very accurately measured.
Here we  use a parametrisation 
\begin{equation}
B = 0.60 \left(\frac{14}{(Q^{2} + M_{V}^{2})^{0.26}} + 1\right) 
\end{equation}
obtained from a fit to experimental data by Mellado \cite{mellado},
and used in their analysis of the predictions of the GW dipole model  
by Caldwell and Soares \cite{cs:01}.
The resulting values 
for  $B$ are shown in Figure \ref{fig:slope}, where we also show the results of
an alternative parameterisation of Kreisel \cite{kreisel:02} to illustrate
the range of values that can be obtained from different fits to the data.
 When comparing the predictions of (\ref{sigmatot}) for the vector meson
cross-sections with data,  it is important to bear in mind that this
uncertainty in the input value of the slope parameter can easily introduce 
errors  up to of order  30$\%$ or so and that, within this range, this error
may be  $Q^2$ dependent.

%\begin{figure}[h]
%\begin{center}
%\includegraphics*[width=8cm,height=8cm]{beta_dgkp_q2_10.eps}\includegraphics*[width=8cm,height=8cm]{beta_dgkp_w_75.eps}
%\caption{\textit{Left}: $W$-dependence of the ratio $\beta$ of the real to imaginary part of the forward amplitude in the FKS model, using the DGKP wavefunctions, for $Q^2 = 10$ GeV$^2$. \textit{Right}: $Q^{2}$-dependence  of $\beta$ in the FKS model, using the DGKP wavefunctions, for $W=75$ GeV.}
%\label{fig:beta_dgkp_q2}
%\end{center}
%\end{figure}

Finally, on comparing  with experimental data, we show
$\sigma^{\mbox{\tiny{TOT}}}=\sigma^{T} + \epsilon \sigma^{L}$ with $\epsilon =
0.98$ in all our plots, although the HERA data range from 0.96 to 1.00.

\section{Results}

In this section we will compare the predictions of the three dipole models
with the experimental data for our three choices of vector meson 
wavefunction, without any adjustment of parameters.  
Before doing so, however, we emphasize again that the uncertainty in 
the slope parameter 
can give rise to $Q^2$-dependent  errors of up to 30$\%$ in the cross-section,
which should be borne in mind when comparing with experiment. This effect
should, hopefully, be much smaller in the ratio $R$. However, it ought not 
to be completely  negligible since the longitudinal and transverse 
cross-sections will presumably have slightly different slope parameters, 
as to some degree they explore dipoles of different sizes.

\subsection{$\rho$ meson production}

\begin{figure}[htbp]
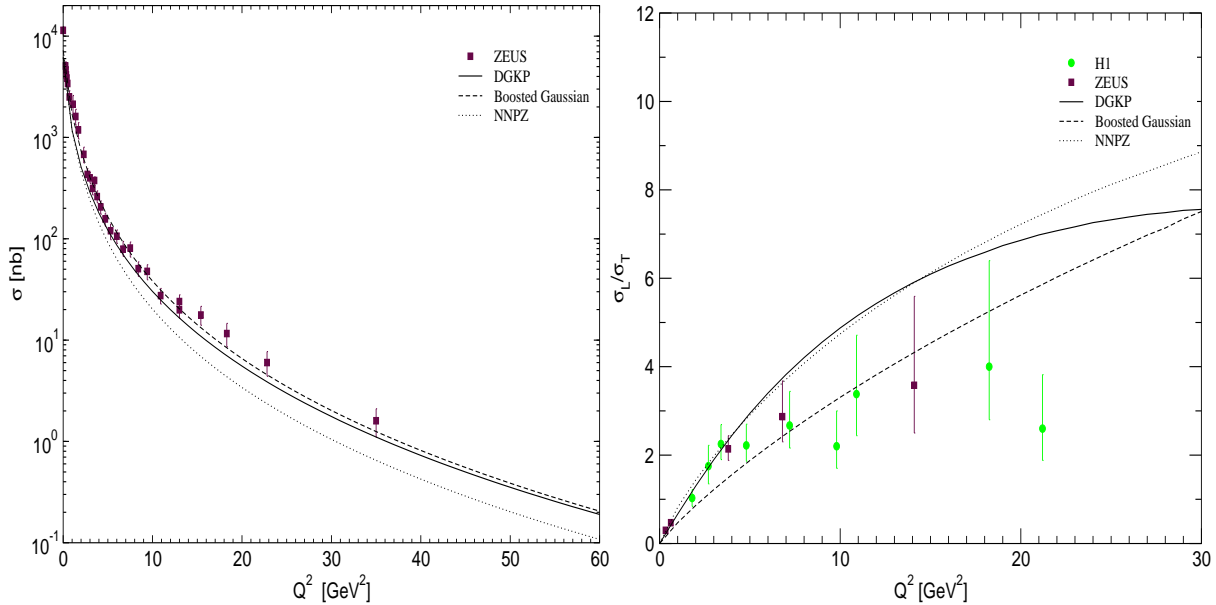

\includegraphics*[width=8cm,height=8cm]{FKS_rho_q2.eps}\includegraphics*[width=8cm,height=8cm]{FKS_rho_R.eps}
\caption{ The $Q^{2}$ dependence of \textit{left} the total cross-section and
\textit{right}  the longitudinal to transverse cross-section ratio for 
$\rho$ production at $W = 75$ GeV in the FKS model using the three different 
meson wavefunctions.  Data from \textit{left} \cite{ZEUS-rho-q2-W-R} and
\textit{right} \cite{ZEUS-rho-q2-W-R, H1-rho-R}.}
\label{FKS_rho_q2_R.eps}
\end{figure}

\begin{figure}[htbp]
\includegraphics*[width=8cm,height=8cm]{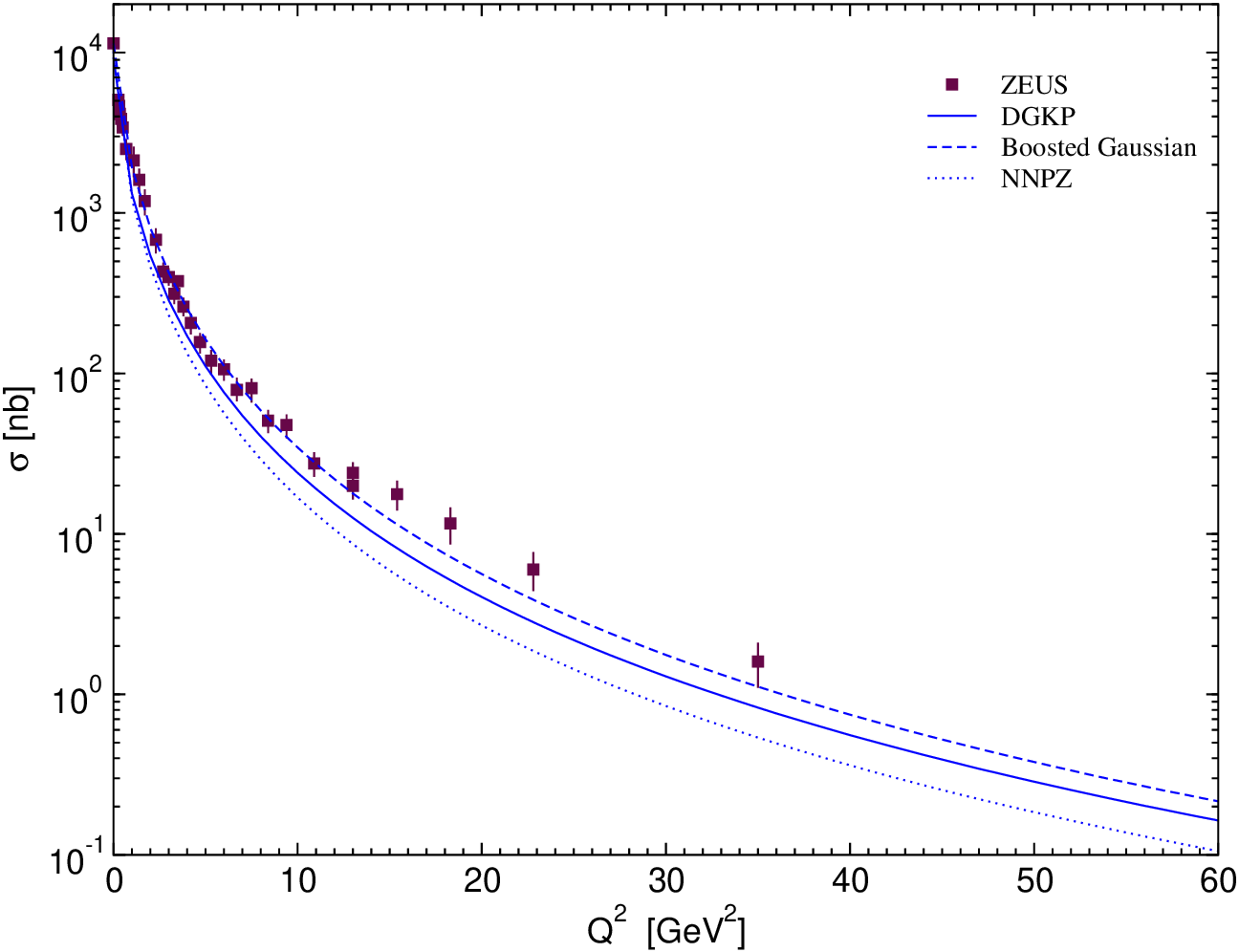}\includegraphics*[width=8cm,height=8cm]{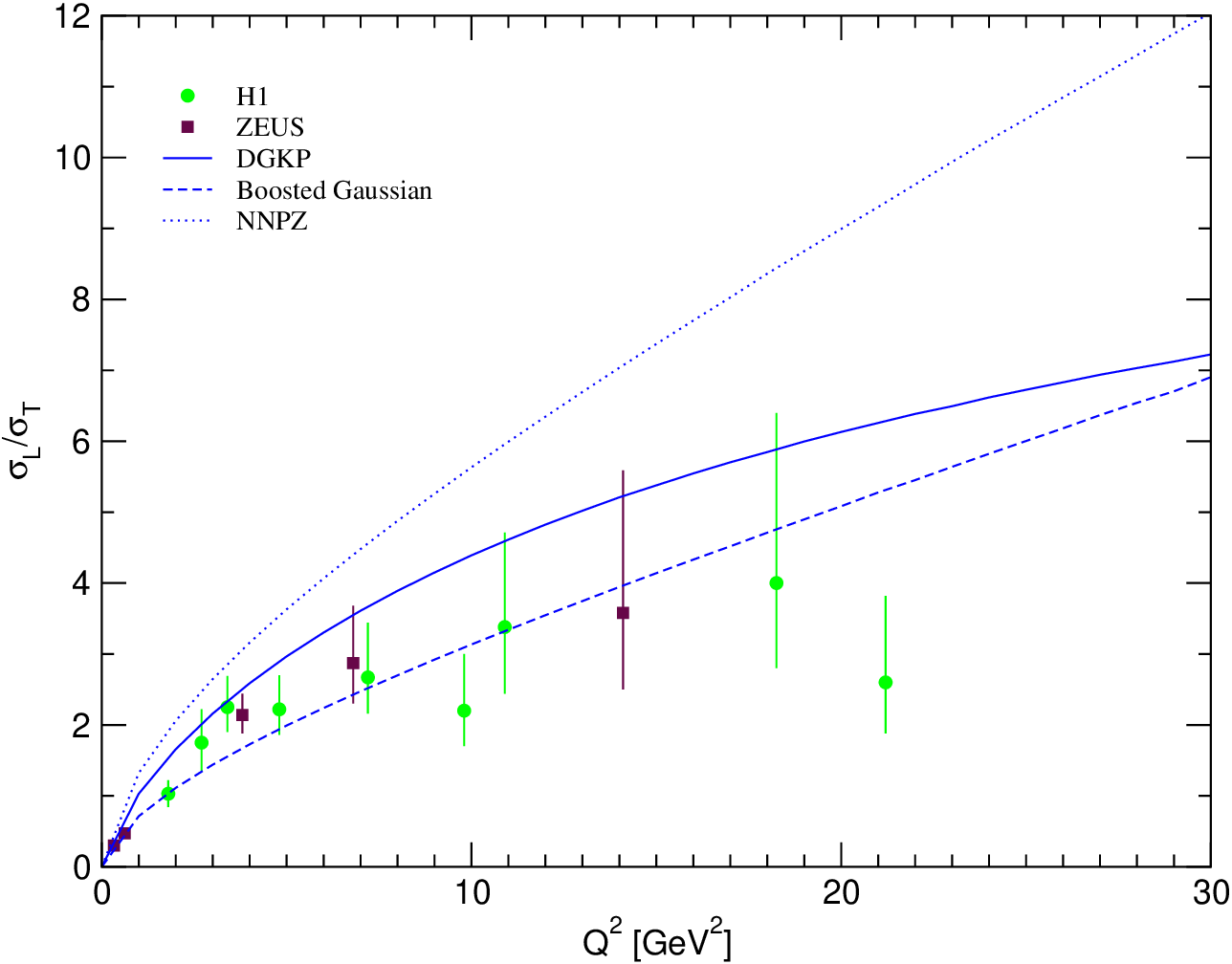}
\caption{  The $Q^{2}$ dependence of \textit{left} the total cross-section and
\textit{right}  the longitudinal to transverse cross-section ratio 
for $\rho$ production at $W = 75$ GeV in the GW model using the three 
different meson wavefunctions.
Data from \textit{left} \cite{ZEUS-rho-q2-W-R} and
 \textit{right} \cite{ZEUS-rho-q2-W-R, H1-rho-R}.}
\label{GW_rho_q2_R.eps}
\end{figure}

\begin{figure}[htbp]
\includegraphics*[width=8cm,height=8cm]{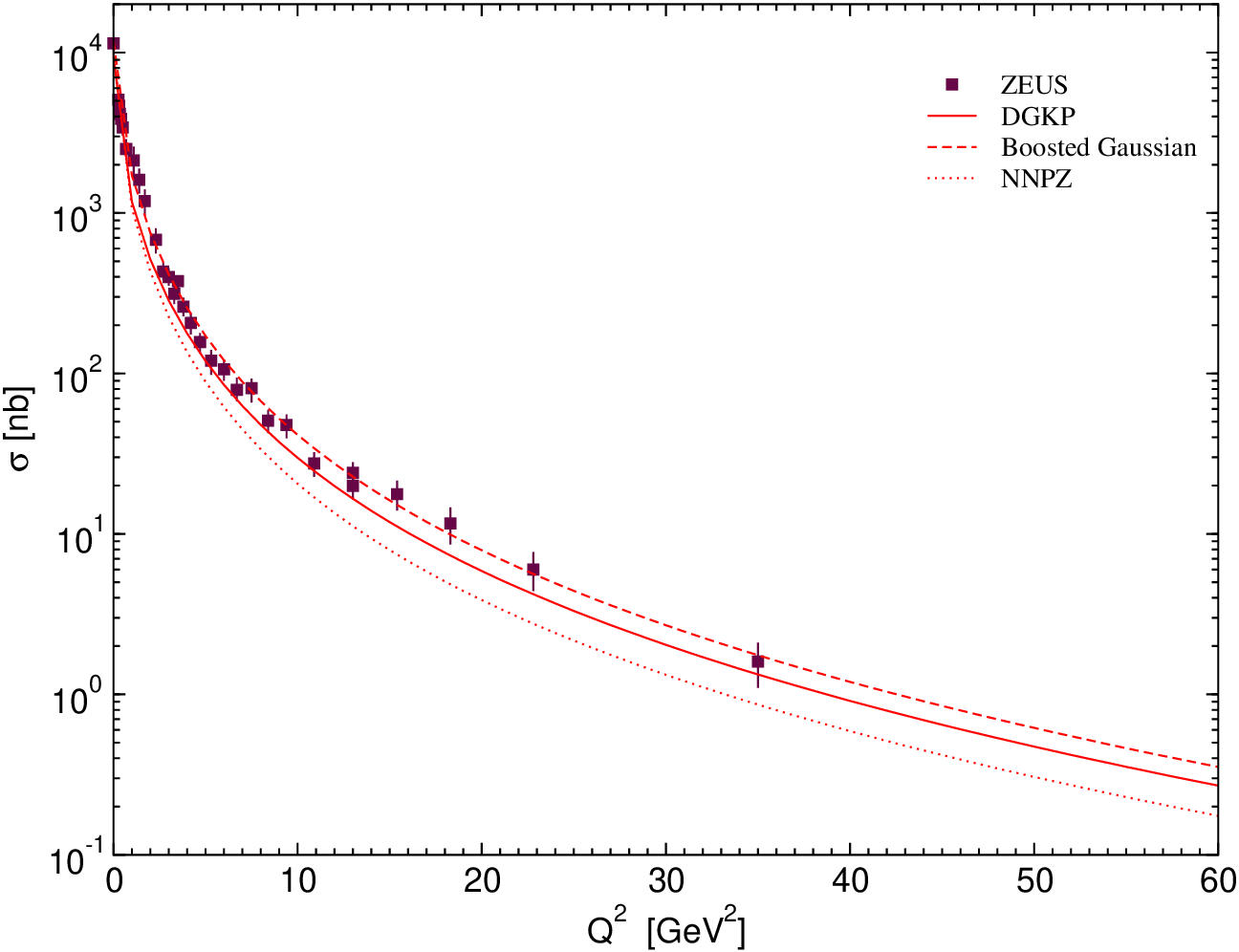}\includegraphics*[width=8cm,height=8cm]{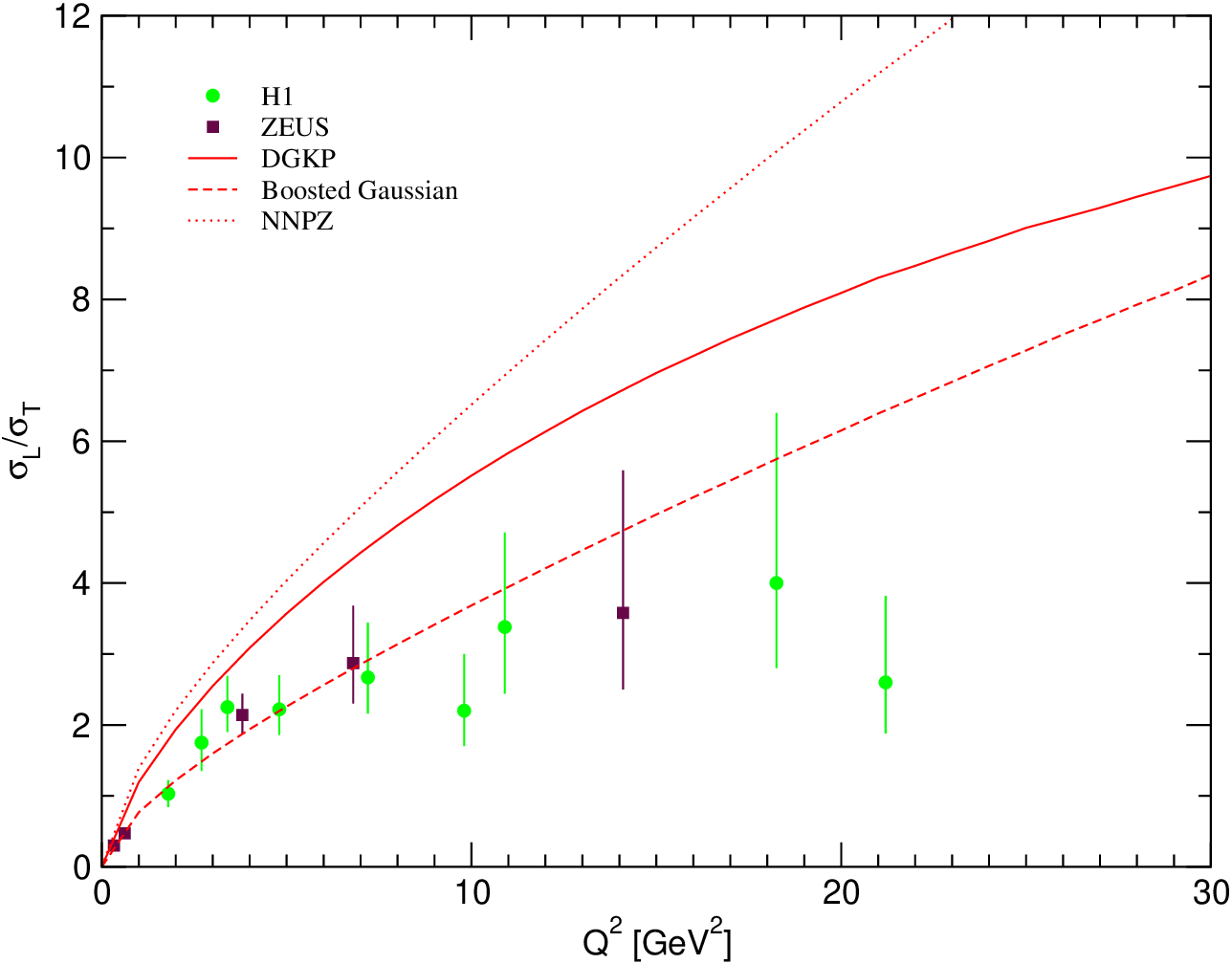}
\caption{  The $Q^{2}$ dependence of \textit{left} the total cross-section and
\textit{right}  the longitudinal to transverse cross-section ratio 
for $\rho$ production at $W = 75$ GeV in the CGC model using the three 
different meson wavefunctions.
Data from \textit{left} \cite{ZEUS-rho-q2-W-R} and
 \textit{right} \cite{ZEUS-rho-q2-W-R, H1-rho-R}.}
\label{CGC_rho_q2_R.eps}
\end{figure}

The predictions of the FKS model for the $Q^2$ dependence of the 
cross-section and for the longitudinal to transverse ratio $R$ are shown 
in Figure \ref{FKS_rho_q2_R.eps} for the three 
different wavefunctions considered. The equivalent plots for the GW
and CGC model are in Figures \ref{GW_rho_q2_R.eps} and \ref{CGC_rho_q2_R.eps}
respectively. 

The ZEUS total cross-section data are at the following centre-of-mass
energies: 0.27 GeV$^2 < Q^2 < $ 0.69 GeV$^2$, $W = 51.1$ GeV, otherwise 
$W=75$ GeV. The H1 data on the longitudinal to transverse ratio are, 
for $Q^2 = 9.8,~18.25$ GeV$^2$, in the range 40 GeV $< W < $ 140 GeV, 
otherwise they are at $W = 75$ GeV. Similarly the ZEUS data are, 
for $Q^2 = 0.33$ GeV$^2$, at $W = 67$ GeV or for $Q^2 = 0.62$ GeV$^2$ at 
$W=47$ GeV. For the theory curves we always take $W=75$ GeV.

For all three dipole models, the
data favour the boosted Gaussian wavefunction, whilst the DGKP wavefunction
produces reasonable agreement for FKS and is rather less satisfactory for
the GW and CGC models. The NNPZ wavefunction is well below the total 
cross-section data for all three  models and, for the GW and CGC models,
in disagreement with the data on the longitudinal to transverse ratio.
However, as noted in section 3.2, although the spurious singularity in this
wavefunction does not  contribute directly to the predicted cross-sections,
it influences them indirectly because it influences the value of the radial 
parameter $R$ deduced from the decay width. In what follows, we will 
therefore focus on the DGKP and boosted Gaussian wavefunctions.

%\textcolor{blue}{Ruben to provide W-dependence figure for CGC. Will need to
%slightly modify next paragraph when I see it. - Graham}

\begin{figure}[htbp]
\includegraphics*[width=8cm,height=8cm]{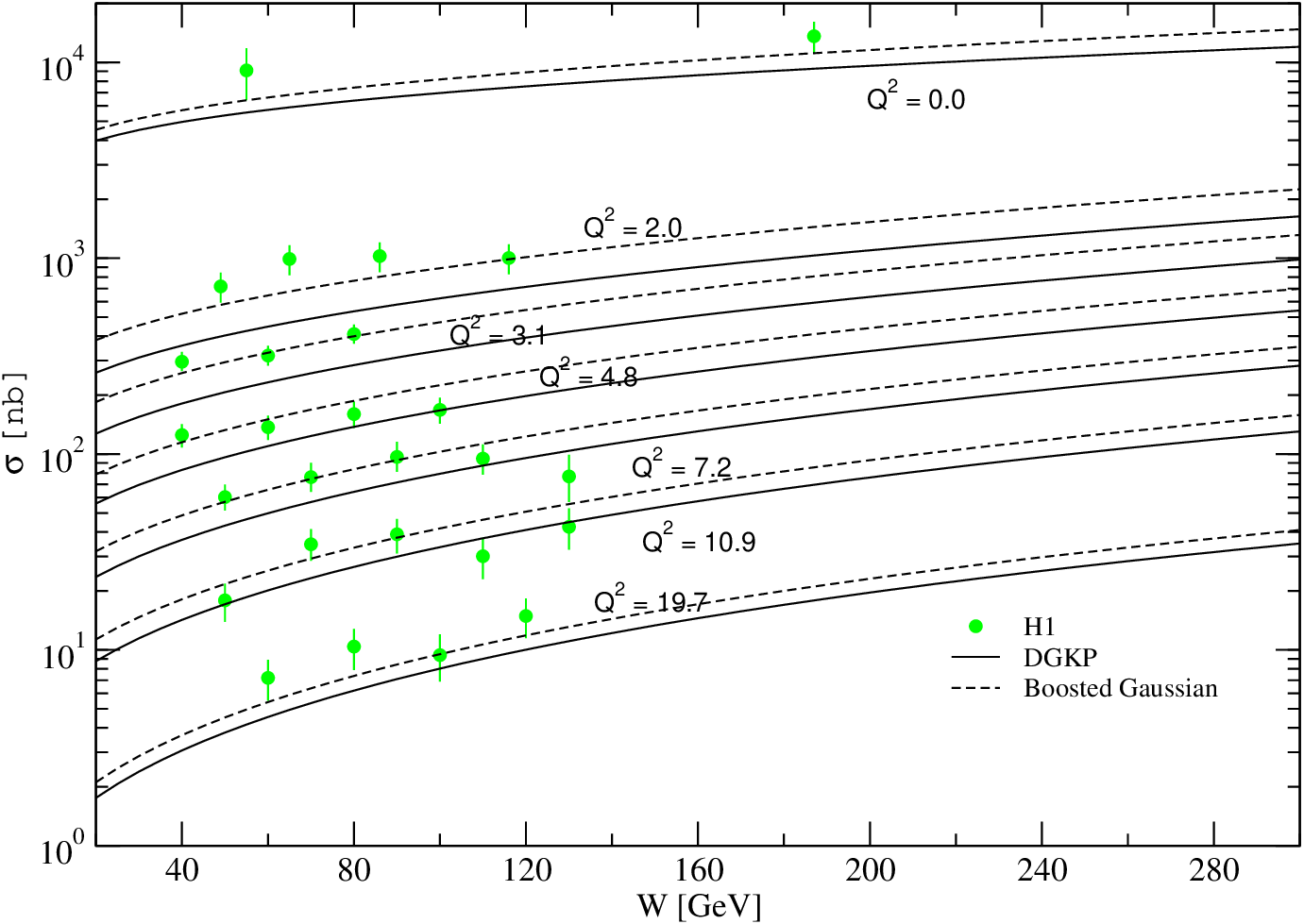}\includegraphics*[width=8cm,height=8cm]{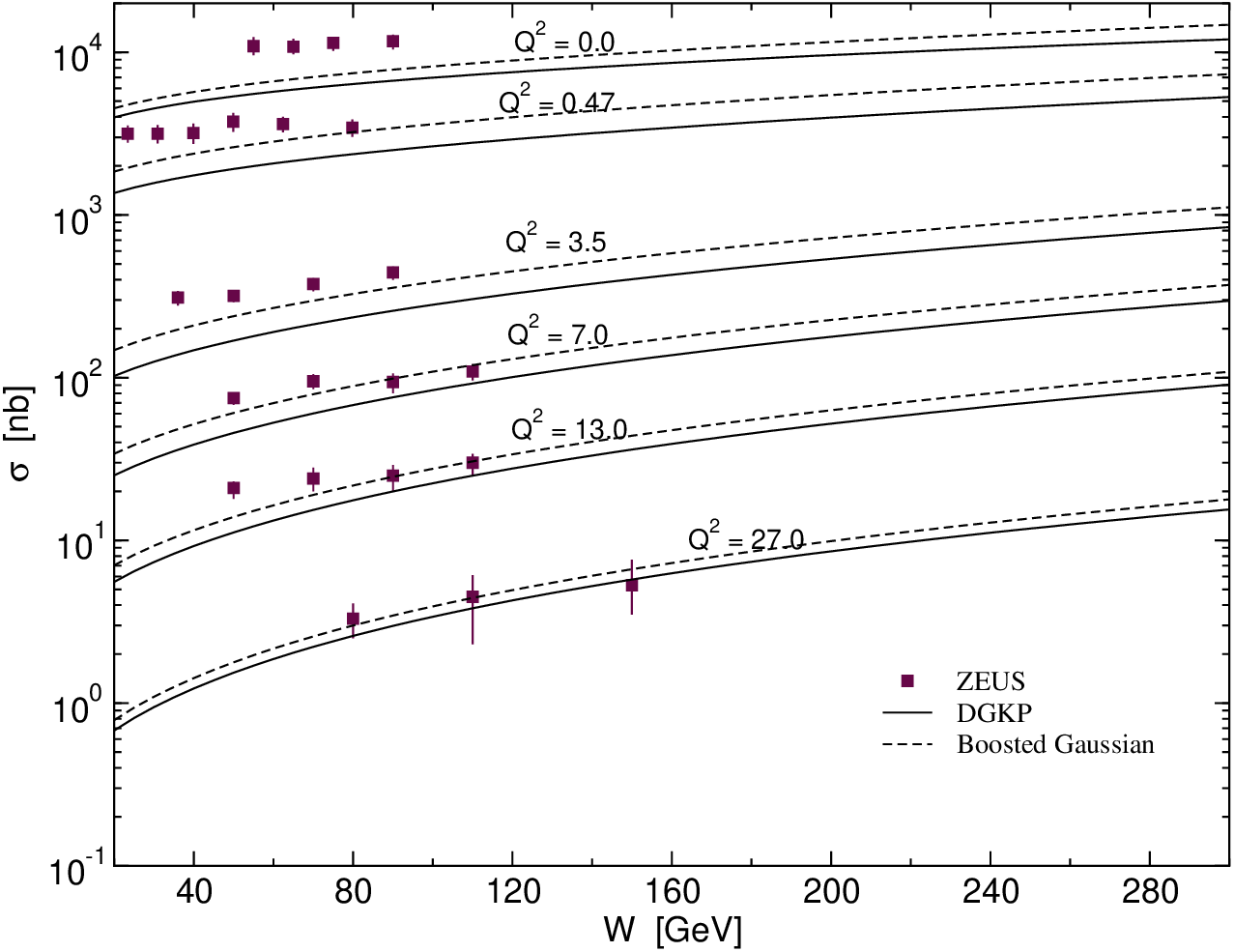}
\caption{The $W$ dependence of the total cross-section for $\rho$ production
at various values of
$Q^2$. 
We use the FKS dipole model and compare the boosted Gaussian and DGKP wavefunctions.  
Data from \textit{left} \cite{H1-rho-W-R} and \textit{right} 
\cite{ZEUS-rho-q2-W-R}.
}
\label{W_rho_FKS.eps}
\end{figure}

\begin{figure}[htbp]
\includegraphics*[width=8cm,height=8cm]{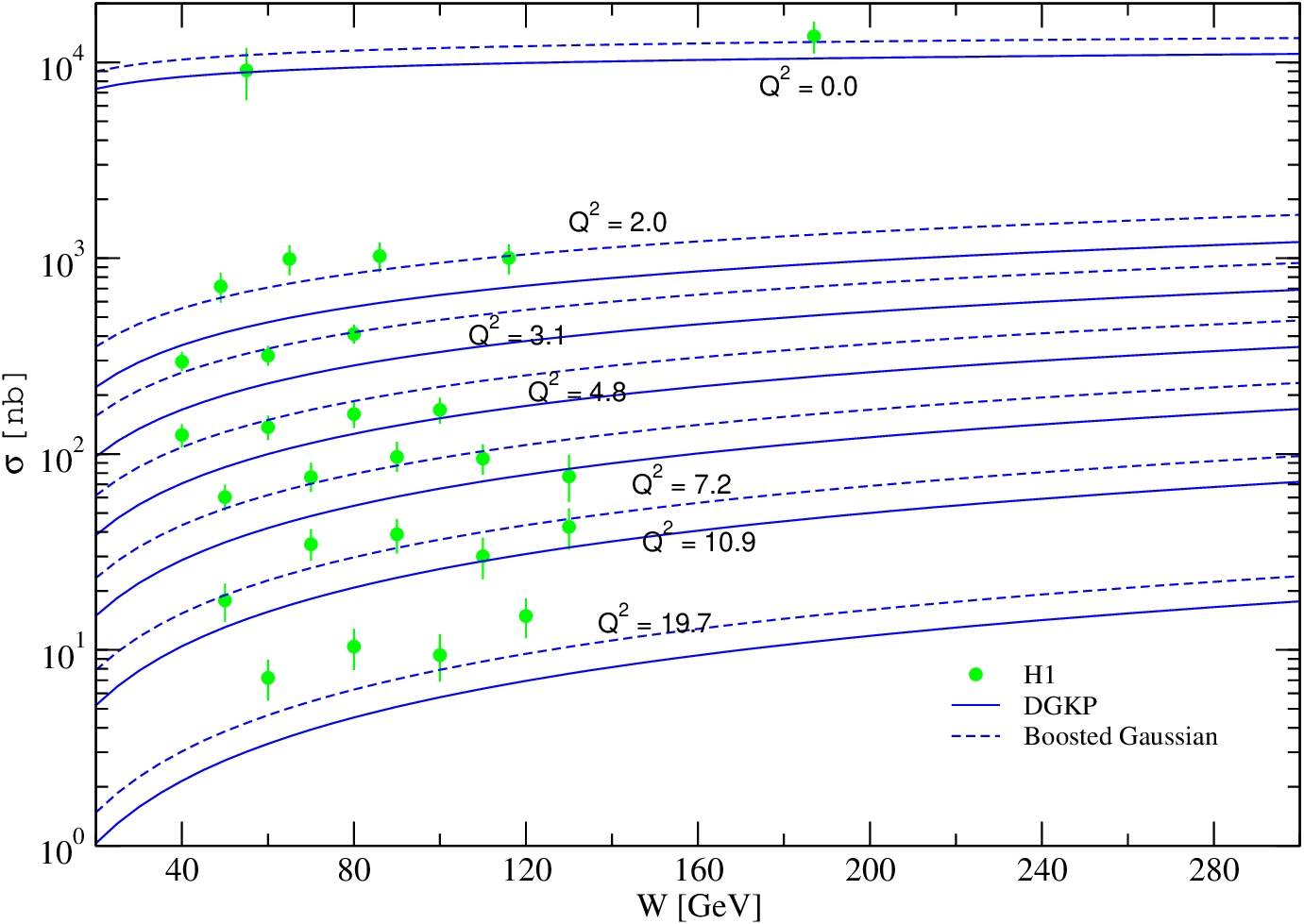}\includegraphics*[width=8cm,height=8cm]{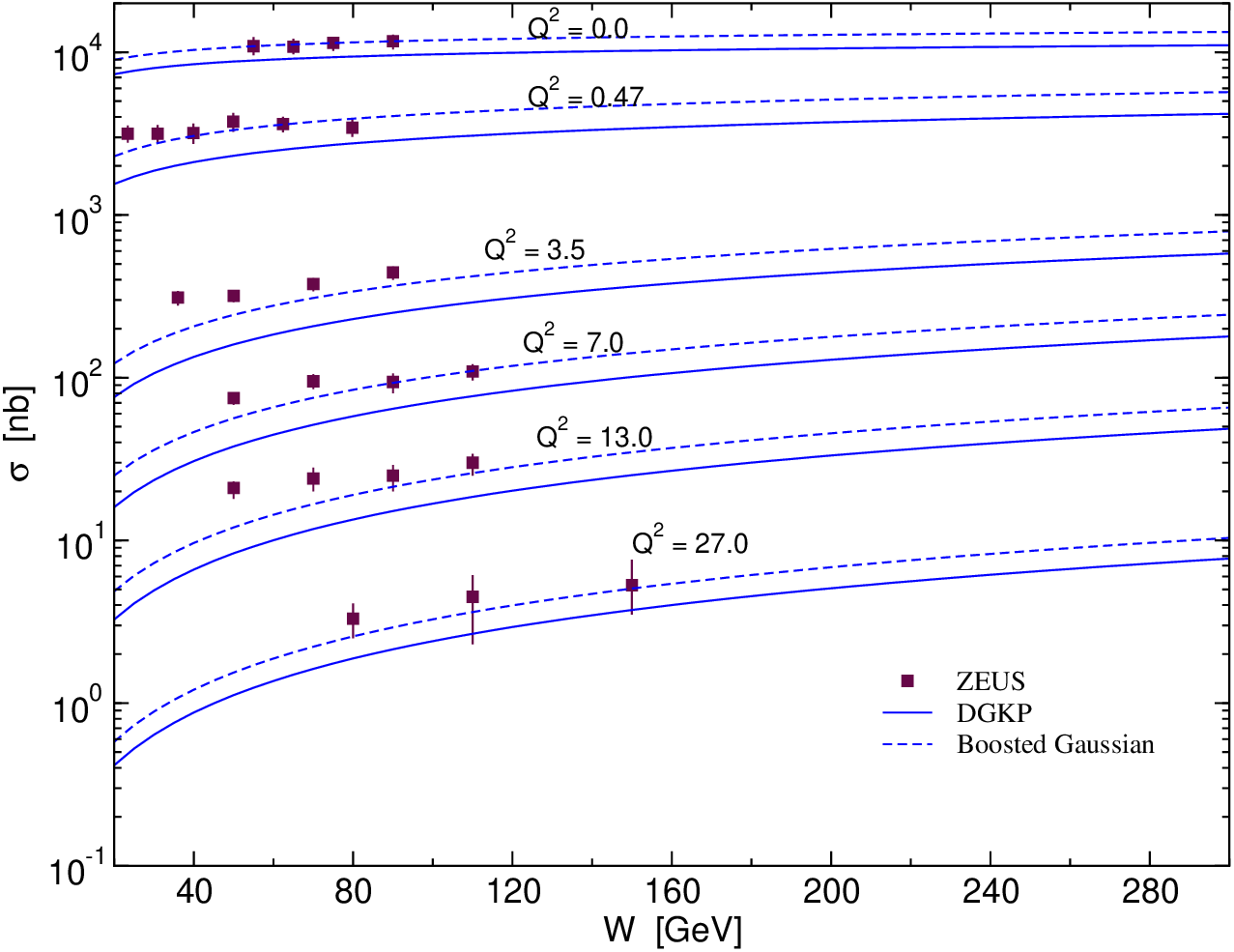}
\caption{The $W$ dependence of the total cross-section for $\rho$ production 
at various values of
$Q^2$. We use the GW dipole model and compare the boosted Gaussian
and DGKP wavefunctions.
Data from \textit{left} \cite{H1-rho-W-R} and \textit{right} 
\cite{ZEUS-rho-q2-W-R}.
}
\label{W_rho_GW.eps}
\end{figure}

\begin{figure}[htbp]
\includegraphics*[width=8cm,height=8cm]{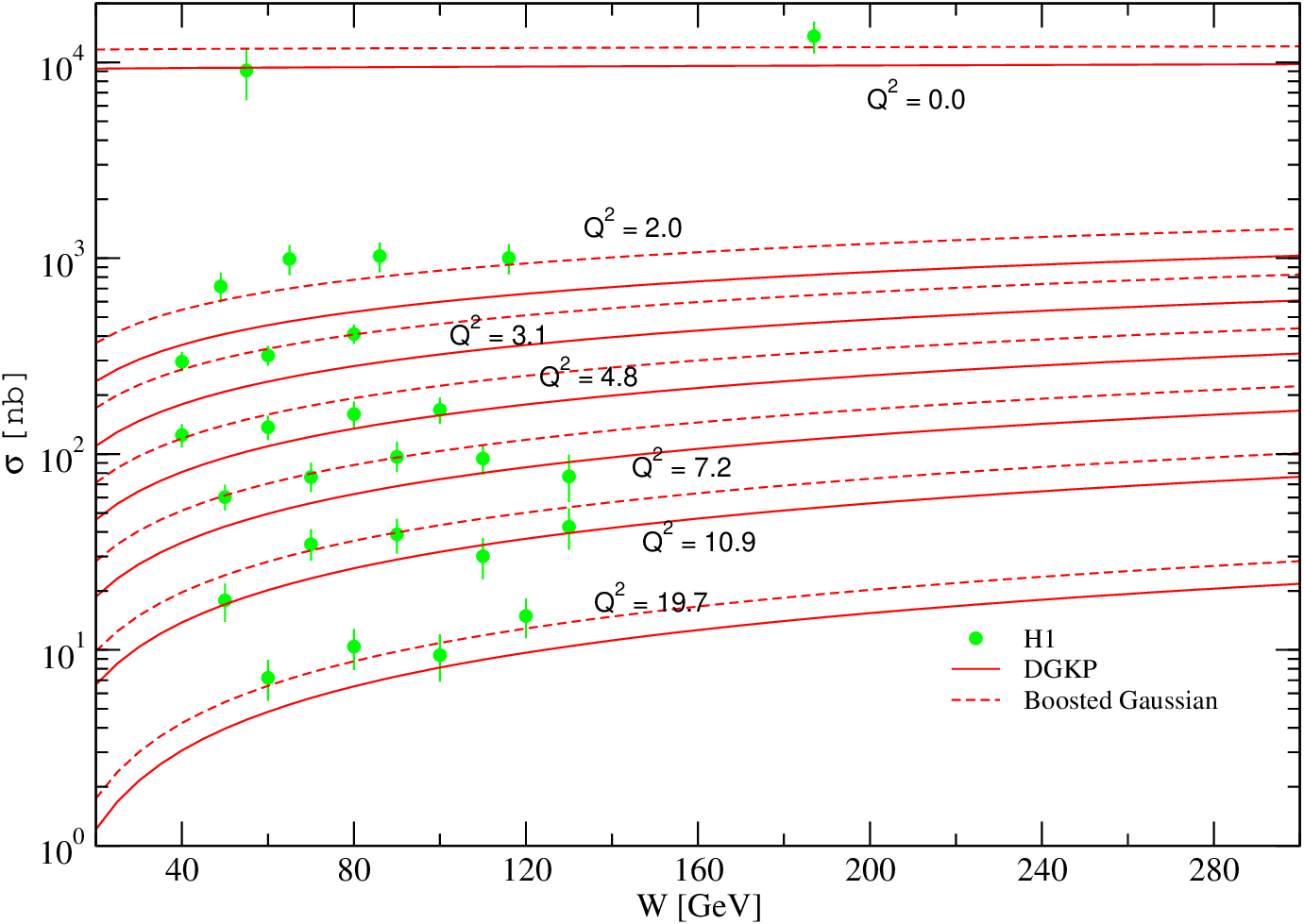}\includegraphics*[width=8cm,height=8cm]{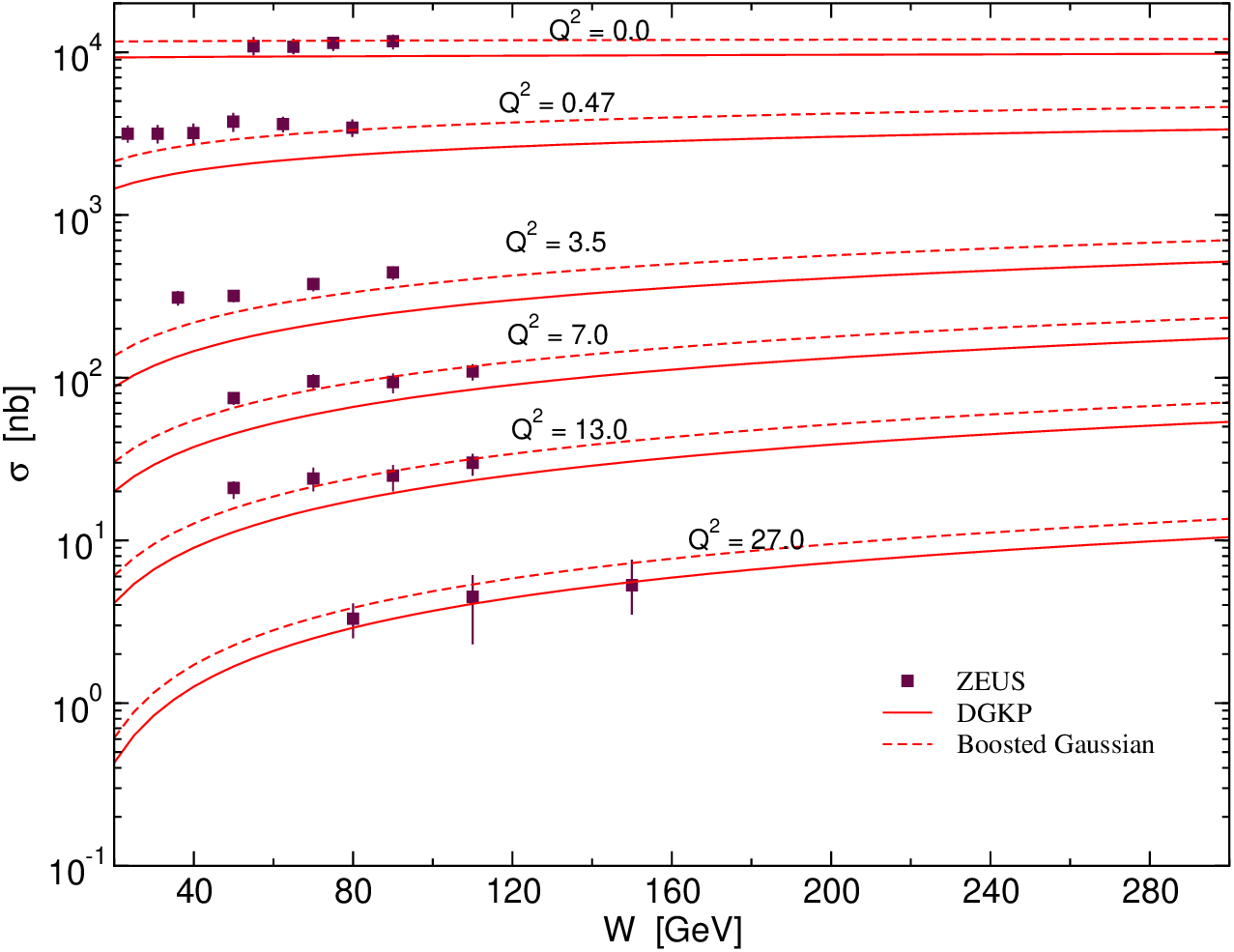}
\caption{The $W$ dependence of the total cross-section for $\rho$ production 
at various values of
$Q^2$. We use the CGC dipole model and compare the boosted Gaussian
and DGKP wavefunctions.
Data from \textit{left} \cite{H1-rho-W-R} and \textit{right} 
\cite{ZEUS-rho-q2-W-R}.
}
\label{W_rho_CGC.eps}
\end{figure}

Figures \ref{W_rho_FKS.eps}, \ref{W_rho_GW.eps} and \ref{W_rho_CGC.eps} show the $W$
dependence at fixed values of $Q^2$ for the $\rho$ meson. Apart from the
normalisation, the $W$-dependence is good in all cases. The normalisation is
best described by the boosted Gaussian wavefunction. With this
wavefunction, the FKS model is in reasonably good agreement with the 
data, except for the ZEUS data at very low $Q^2$; while the    
GW and CGC models give reasonable agreement everywhere.

This last comment contrasts somewhat with the work of 
 Caldwell and Soares \cite{cs:01}, who have already presented predictions
 for the GW  model using a boosted Gaussian wavefunction. These authors
also  found good agreement with the data on the  longitudinal to transverse
ratio and with the $W$ dependence at fixed $Q^2$ apart from the normalisation.
However their results for the normalisation, and hence, implicitly,
for the $Q^2$-dependence of the production cross-section, were relatively
poor. However, these authors did not implement the leptonic decay width 
constraint but fixed the radial parameter $R$ in 
the boosted Gaussian by requiring that the exponential in $R$ of the 
wavefunction gives a value of $1/e$ when the $q\bar{q}$ invariant mass is 
equal to the meson's mass. This yielded $R^{2}= 15.5~\mbox{GeV}^{-2}$ 
for $\rho$ and $R^{2} = 8.3~\mbox{GeV}^{-2}$ for $\phi$ compared to our 
values shown in Table 4. In addition, they neglected real parts, which
can result in a 20$\%$ reduction in the cross-section for large $Q^2$.

%%%%% FIGURES FOR RHO

\subsection{$\phi$ meson production}
\begin{figure}[htbp]
\includegraphics*[width=8cm,height=8cm]{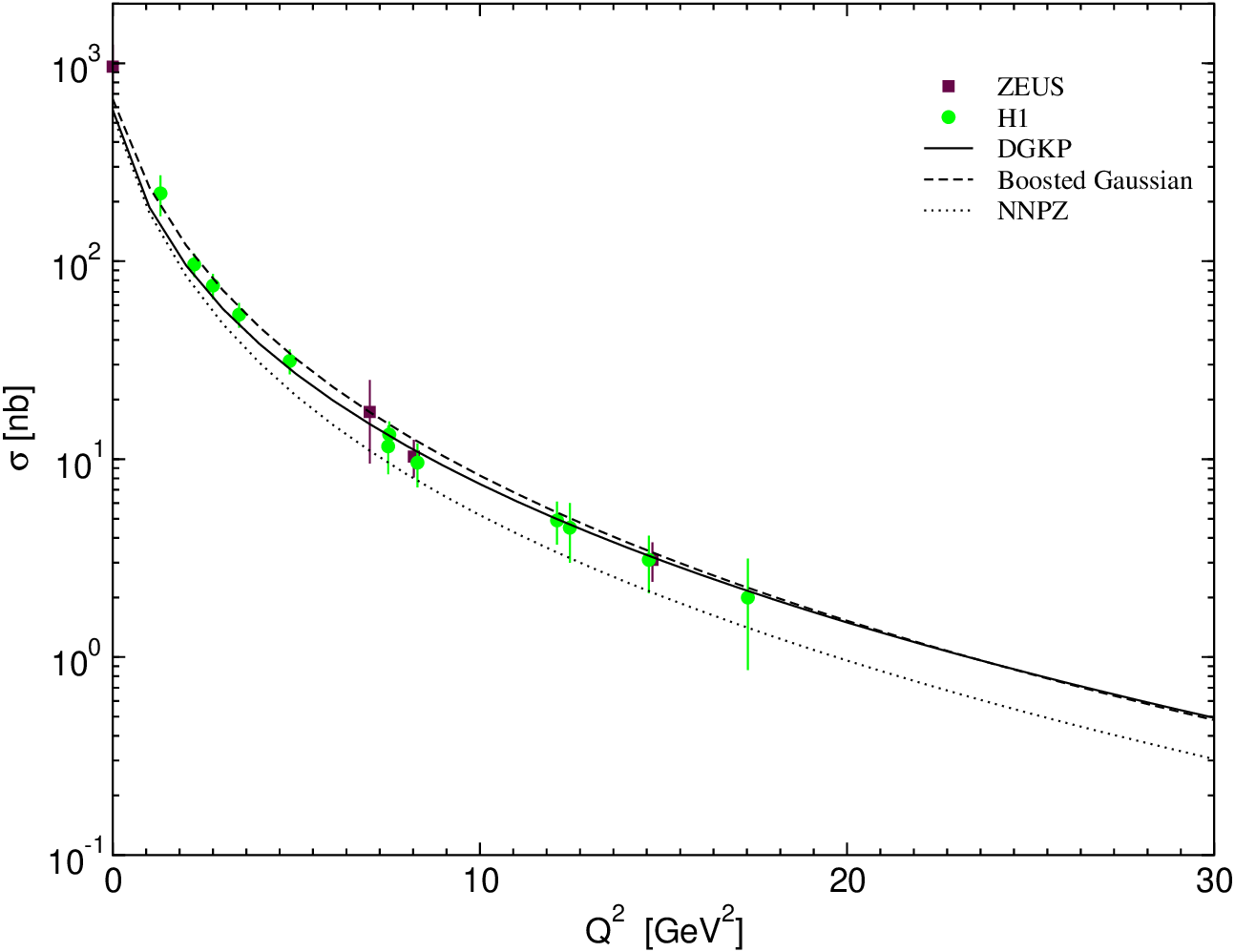}
\includegraphics*[width=8cm,height=8cm]{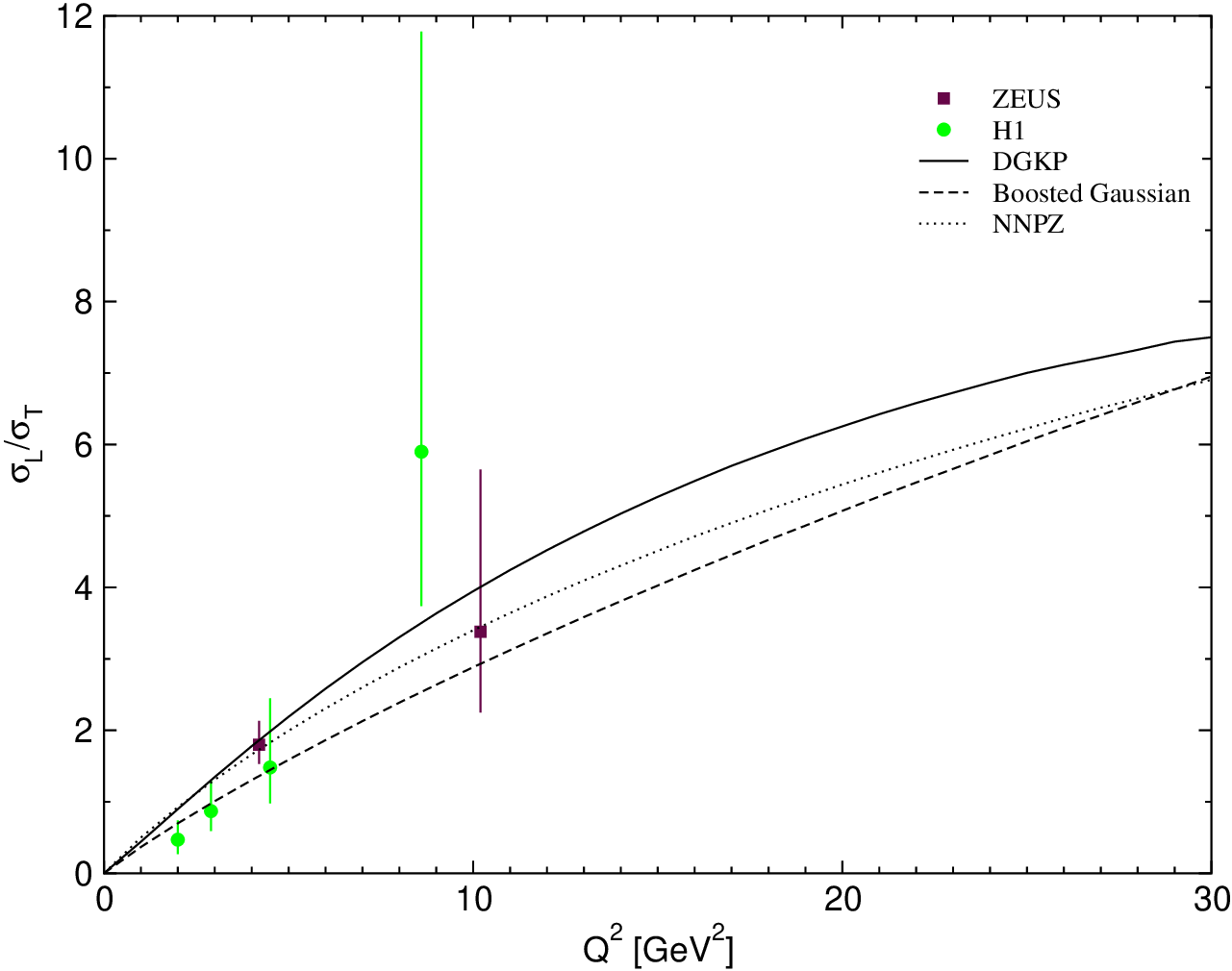}
\caption{ The $Q^{2}$ dependence of \textit{left} the total cross-section 
and \textit{right} the longitudinal to transverse cross-section ratio for
$\phi$ production  at $W=90$ GeV in the FKS model using
the three different meson wavefunctions.
Data from  \textit{left} \cite{H1-phi-q2,ZEUS-phi-q2-W} and \textit{right}
 \cite{H1-phi-R,ZEUS-phi-R}.}
\label{FKS_phi_q2_R.eps}
\end{figure}

The corresponding plots to those in the previous section are repeated for
the $\phi$ meson in Figures \ref{FKS_phi_q2_R.eps}, \ref{GW_phi_q2_R.eps} and \ref{CGC_phi_q2_R.eps}. 
The H1 total cross-section data are at the following centre-of-mass
energies: for $Q^2=7.5,8.3,12.5,14.6,17.3$ GeV$^2$, $W = 100$ GeV, 
otherwise $W = 75$ GeV. Similarly for the ZEUS data, for $Q^2=0.0, 8.2, 
14.7$ GeV$^2$, $W= 70,94,99$ GeV respectively.  
The H1 data on the longitudinal to transverse ratio are
in the range 40 GeV $< W < $ 130 GeV. Similarly the ZEUS data are in the
range 25 GeV $< W < 120$ GeV. For the theory curves we always take
$W=90$ GeV.

The situation is rather similar, as one might
expect, to that of the $\rho$, i.e. all three dipole models  tend to do 
rather well
with either of the DGKP or boosted Gaussian wavefunctions, while the NNPZ
wavefunction is less  satisfactory.

% FIGURES FOR PHI PRODUCTION

\begin{figure}[htbp]
\includegraphics*[width=8cm,height=8cm]{GW_phi_q2.eps}
\includegraphics*[width=8cm,height=8cm]{GW_phi_R.eps}
\caption{  The $Q^{2}$ dependence of \textit{left} the total cross-section 
and \textit{right} the longitudinal to transverse cross-section ratio for
$\phi$ production  at $W=90$ GeV in the GW model using the three
different wavefunctions.
Data from  \textit{left} \cite{H1-phi-q2,ZEUS-phi-q2-W} and \textit{right}
\cite{H1-phi-R,ZEUS-phi-R}.
}
\label{GW_phi_q2_R.eps}
\end{figure}

\begin{figure}[htbp]
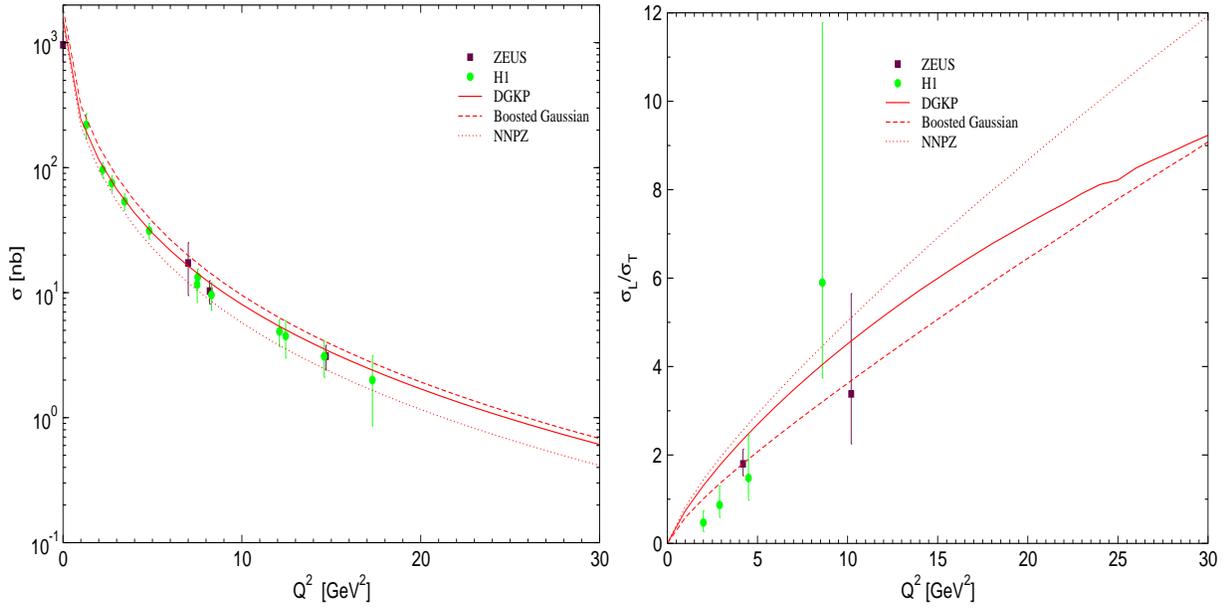

\includegraphics*[width=8cm,height=8cm]{CGC_phi_q2.eps}
\includegraphics*[width=8cm,height=8cm]{CGC_phi_R.eps}
\caption{  The $Q^{2}$ dependence of \textit{left} the total cross-section 
and \textit{right} the longitudinal to transverse cross-section ratio for
$\phi$ production  at $W=90$ GeV in the CGC model using the three
different wavefunctions.
Data from  \textit{left} \cite{H1-phi-q2,ZEUS-phi-q2-W} and \textit{right}
\cite{H1-phi-R,ZEUS-phi-R}.
}
\label{CGC_phi_q2_R.eps}
\end{figure}

\section{Conclusions}

We have performed a detailed study, comparing the predictions on
$\rho$ and $\phi$ meson electroproduction arising from three different
models of the meson wavefunction in combination with three different
models for the fundamental dipole cross-section. Our results are broadly
encouraging and support the use of the dipole approach. 

The data  can be explained rather well using the dipole model
of Forshaw, Kerley and Shaw, or those  of Golec-Biernat and W\"usthoff and of
Iancu, Itakura and Munier,  in
conjunction with either the boosted Gaussian or DGKP meson
wavefunction. Certainly, we anticipate that excellent agreement could be
obtained if one decided to tune the meson wavefunctions. Note that 
agreement extends to the ratio of longitudinal to transverse meson 
production. The NNPZ wavefunction, which has, as noted, an unphysical
singularity at $z = 0.5$, $r = 0$, is not so successful. 

For the future, it is clear that one could use the high quality data 
from HERA to constrain the meson wavefunctions provided the dipole
cross-section is sufficiently constrained. 

\section*{Acknowledgements}
This research was supported in part by the UK Particle Physics and Astronomy 
Research Council.
R.S. was partly funded by a University of Manchester research scholarship. 
We should like to thank S. Munier, M. Soares and A. Stasto for useful 
discussions. We also thank M. V. T. Machado for useful comments on the text.


\begin{thebibliography}{100}

% FKS dipole
\bibitem{fks1} J.~R.~Forshaw, G.~Kerley and G.~Shaw, Phys. Rev. {\bf D60}
(1999) 074012.

\bibitem{fks2} J.~R.~Forshaw, G.~Kerley and G.~Shaw, Nucl. Phys. {\bf A675} (2000) 80c.

%DVCS-dipole
\bibitem{mss:02} M.~McDermott, R.~Sandapen and G.~Shaw, Eur. Phys. J. C. {\bf 22} (2002) 655.

%F2D3:H1 and ZEUS data
\bibitem{f2d3dat}   C.~Adloff {\it et al.}, H1 Collab.,  Z. Phys. {\bf  C76}
(1997) 613; J.~Breitweg {\it et al.}, ZEUS Collab., Eur. Phys. J. {\bf  C6}
(1999) 43. 

%DVCS: H1 and ZEUS data
\bibitem{DVCSdat} C.~Adloff {\it et al.}, H1 Collab., Phys.~Lett.~{\bf B517}
(2001) 47;  S.~Chekanov {\it et al.}, Phys. Lett. {\bf B573} (2003) 46.

%GW dipole
\bibitem{gw:99a} K.~Golec-Biernat and M.~W\"{u}sthoff, Phys. Rev. {\bf D59} (1999)014017.

\bibitem{gw:99b} K.~Golec-Biernat and M.~W\"{u}sthoff, Phys. Rev. {\bf D60} (1999) 114023.

% CGC dipole
\bibitem{iim03} E.~Iancu, K.~Itakura and S. Munier, hep -ph/0310338.  

%DGKP wfn
\bibitem{dgkp:97} H.~G.~Dosch, T.~Gousset, G.~Kulzinger and H.~J.~Pirner, Phys. Rev. \textbf{D55} (1997) 2602.

%NNPZ wfn
\bibitem{nnpz:97} J.~Nemchik, N.~N.~Nikolaev, E.~Predazzi and B.~G.~Zakharov,  Z.Phys. \textbf{C75} (1997) 71.

%Dipole original refs
\bibitem{dipole}  N.~N.~Nikolaev and B.~G.~Zakharov, Z. Phys. {\bf  C49}
(1991) 607; Z. Phys. {\bf C53} (1992) 331; A.~H.~Mueller, Nucl. Phys. 
{\bf B415} (1994) 373; A.~H.~Mueller and B.~Patel, Nucl. Phys. {\bf B425} 
(1994) 471.

%Presentation of various dipole models with refs.
\bibitem{amirim} M.~McDermott, \textit{The dipole picture of small $x$ physics (A summary of the Amirim meeting)}, hep-ph/0008260v2.

\bibitem{FGS}  L.~Frankfurt, V.~Guzey and M.~Strikman,  Phys. Rev. {\bf D58} (1998) 094039.

%Generalized meson  dominance
\bibitem{GVD1}  H.~Fraas, B.~J.~Read and D.~Schildknecht, Nucl. Phys. 
{\bf B86} (1975) 346.

\bibitem{GVD2} G.~Shaw,  Phys. Rev. {\bf D47} (1993) R3676; 
G.~Shaw, Phys. Lett. {\bf B228} (1989) 125; 
 P.~Ditsas and G.~Shaw, Nucl.Phys. {\bf B113} (1976) 246.
 
\bibitem{GVD3}  G.~Cvetic, D.~Schildknecht, B.~Surrow and M.~Tentyukov,
Eur. Phys. J. {\bf C20} (2001) 77.

%Saturation
\bibitem{levin} L.V.~Gribov, E.M.~Levin and M.G.~Ryskin, Phys. Rep. {\bf 100}
 (1983) 1.

%FKS:saturation
\bibitem{fks3} J.~R.~Forshaw, G.~Kerley, G.~Shaw, Proc 8th Int. Workshop on 
Deep Inelastic Scattering, Eds. J.~A.~Gracey and T.~Greenshaw, World Scientific, 2001. hep-ph/0007257.

% QCD evolution in GW
\bibitem{gw:02} J.~Bartels, K.~Golec-Biernat and H.~Kowalski, Phys. Rev {\bf D66} (2002) 014001. 


%LO pQCD dipole cross-section
%\textcolor{blue}{Cite \cite{frs:97} only if keeping first sentence in Section 2.3; otherwise comment out}
%\bibitem{frs:97} L.~Frankfurt, A.~Radyushkin and M.~Strikman, Phys. Rev. {\bf{D55}} (1997) 98.

%Light-cone formalism and wfns
\bibitem{bl:80} S.~J.~Brodsky and G.~P.~Lepage, Phys. Rev. \textbf{D22} (1980)
2157.

%Gauge-invariance issues in VMP
\bibitem{hl:98} A.~Hebecker and P.~V.~Landshoff, Phys. Lett. {\bf{B419}} (1998) 393.

%Constraints on meson wfns written explicitly in there.
\bibitem{munieretal:01} S.~Munier, A.~Mueller and A.~Stasto, Nucl. Phys. \textbf{B603} (2001) 427.

\bibitem{halperin-zhitnitsky:97} H.~Halperin and A.~Zhitnitsky, Phys. Rev. \textbf{D56} (1997) 184.

% z-dependence of DGKP wfn
\bibitem{wsb:85} M.~Wirbel, B.~Stech and M.~Bauer, Z. Phys. \textbf{C29} (1985) 637.

\bibitem{pdg} Particle data group:  K.~Hagiwara et al., Phys. Rev. D66 (2002) 010001.


%Brodsky-LePage boost
\bibitem{bhl:80} S.~J.~ Brodsky, T.~Huang and G.~P.~LePage, SLAC-PUB-2540 (1980), Shorter version contributed to 20th Int. Conf. on High Energy Physics, Madison, Wisc., Jul 17-23, 1980. 

%Discussion of BL boost; generalization for off-shell quarks
\bibitem{dgs:01} A.~Donnachie, J.~Gravelis and  G.~Shaw,  Phys. Rev. D63 (2001) 114013.

%Running coupling of NNPZ
\bibitem{nnpz:94} J. Nemchik, N. N. Nikolaev and B. G. Zakharov, Phys. Lett. \textbf{B341} (1994) 228.

%MFGS dipole
%\textcolor{blue}{Keep MFGS ref only if footnote on page 21 is kept; otherwise comment out}
%\bibitem{mfgs1} M.~McDermott, L.~Frankfurt, V.~Guzey and M.~Strikman,  Eur. Phys. J. {\bf C16} (2000) 641.


%Diffractive slope parametrization
\bibitem{mellado} B.~Mellado, unpublished result given in \cite{cs:01}.

%GW dipole and VMP
\bibitem{cs:01} A.~Caldwell and M.~Soares, Nucl. Phys. \textbf{A696} (2001) 125.
%B-slope
\bibitem{kreisel:02} A.~Kreisel, paper  presented at LISHEP 02, hep-ex/020801v1.


% RHO
\bibitem{ZEUS-rho-q2-W-R} J.~Breitweg {\it et al.}, ZEUS Collab., Eur. Phys. J. \textbf{C6} (1999) 603.



\bibitem{H1-rho-R} S.~Aid {\it et al.}, H1 Collab., Nucl. Phys. \textbf{B463} (1996) 3.

\bibitem{H1-rho-W-R} C.~Adloff {\it et al.}, H1 Collab., Eur. Phys. J. \textbf{C13} (2000) 371; H1 Collab., Nucl. Phys. \textbf{B468} (1996) 3.

% PHI
\bibitem{H1-phi-q2} C.~ Adloff {\it et al.}, H1 Collab., Phys. Lett. \textbf{B483} (2000) 360; Z. Phys. \textbf{C75} (1997) 607.

\bibitem{ZEUS-phi-q2-W} M.~Derrick {\it et al.}, ZEUS Collab., Phys. Lett. \textbf{377B} (1996) 259; Phys. Lett. \textbf{B380} (1996) 220.

\bibitem{H1-phi-R} C.~Adloff {\it et al.}, H1 Collab., Phys. Lett. \textbf{B483} (2000) 360.

%\textcolor{blue}{Is there a proper journal ref. for ref. below ?}
\bibitem{ZEUS-phi-R} ZEUS Collaboration, submitted paper 793  to ICHEP 98.



\end{thebibliography}
\end{document}